\documentclass[12pt]{article}
\usepackage[a4paper, textheight=280mm, top=10mm, bottom=10mm]{geometry}
\usepackage{subfiles}
\usepackage{upgreek}
\usepackage[utf8]{inputenc}
\usepackage[centertags]{amsmath}
\usepackage{amssymb}
\usepackage{bm}
\usepackage{amsbsy}
\usepackage{graphicx}
\graphicspath{{Images/}}
\usepackage[labelformat = empty]{subcaption}
\usepackage{appendix}
\usepackage[main=english,russian]{babel}
\usepackage{mathrsfs}
\usepackage{epsfig}
\usepackage{color}
\usepackage{euscript}
\usepackage{ulem}
\usepackage{diagbox}
\usepackage{multirow, makecell}
\usepackage{seqsplit}
\usepackage{microtype}
\usepackage{xurl}
\usepackage{jheppubm}
\usepackage{tikz}[[arr/.style={->,thick}]
\tikzstyle{block} = [draw, fill=white, rectangle, 
    minimum height=3em, minimum width=6em]
\tikzstyle{pinstyle} = [pin edge={to-,thin,black}]

\newcommand{\mcA}{\mathcal{A}}

\newcommand{\be}{\begin{equation}}
\newcommand{\ee}{\end{equation}}
\newcommand{\bea}{\begin{eqnarray}}
\newcommand{\eea}{\end{eqnarray}}

\newcommand{\ff}{\mathfrak{f}}

\newcommand{\fg}{\mathfrak{g}}

\newcommand{\cV}{{\cal V}}

\newcommand{\cO}{{\cal O}}
\numberwithin{equation}{section}

\definecolor{dgreen}{rgb}{0,0.6,0}

\definecolor{darkblue}{rgb}{0., 0, 1}

\definecolor{purple}{rgb}{0.65,0.,0.78}

\definecolor{orange}{rgb}{0.89,0.46,0.15}

\definecolor{darkyellow}{rgb}{0.7, 0.6, 0.0}

\title{
Optimizing the dilaton potential in holographic QCD via phase-transition constraints
}
\author{Irina Ya. Aref’eva, Alexander V. Polevov, Pavel S. Slepov}

\affiliation{Steklov Mathematical Institute, Russian Academy of  Sciences, \\ Gubkina str. 8, 119991, Moscow, Russia}

\emailAdd{arefeva@mi-ras.ru}
\emailAdd{avpolevov@mi-ras.ru}
\emailAdd{slepov@mi-ras.ru}

 \abstract{In bottom-up holographic QCD (HQCD), two primary approaches are commonly employed: the potential reconstruction method, where the background geometry is fixed a priori and the dilaton potential is derived, and the direct method, where the potential is specified explicitly. While the reconstruction method is highly effective for capturing the phenomenological properties of QCD under extreme conditions, a subtlety arises in that the reconstructed potential depends on the chosen holographic boundary conditions.
\\

To address this issue, we propose constructing a dilaton potential designed to reproduce the typical features of HQCD phase transitions obtained by the reconstruction method, rather than one derived solely from HQCD vacuum solution.  Using a light-quark model as an example, we construct a  potential that captures the principal features of the phase structure of this model. Its parameters are determined at zero chemical potential by minimizing the deviation between the temperature dependence on the horizon position in the direct and reconstructed models and by fitting to the same position of a critical endpoint (CEP).
\\

The resulting  minimal-potential model  satisfactorily recovers the physics of the light-quark model even at relatively small chemical potential. Specifically, it yields a first-order phase transition (FOPT) line close to those obtained from the reconstruction method. Additionally, the minimal-potential model reproduces a rising Cornell-like quark-antiquark potential up to the hadronic distance $\ell=1.5$ fm, which we use as a phenomenological finite-distance criterion for confinement.
The location of the confinement/deconfinement crossover in this minimal-potential model is also close to its counterpart in the original light-quark model.
 }

 
\begin{document}

\maketitle

\newpage

\section{Introduction}

\qquad Understanding the strongly coupled regime of gauge theories---and QCD in particular---is a central challenge in modern theoretical physics. It is essential for describing the quark-gluon plasma created in ultra-relativistic heavy-ion collisions at facilities such as RHIC, LHC, NICA and FAIR.  First-principles lattice QCD accurately describes the theory at zero baryon chemical potential $\mu = 0$, but it struggles at finite $\mu$ due to the fermion sign problem \cite{Blankenbecler:1981jt}, making direct Monte Carlo simulations unfeasible.

Holographic duality \cite{Maldacena:1997re} provides an alternative approach by relating the strongly coupled regime of a gauge theory to a weakly coupled gravitational theory in one additional dimension. Holographic models are divided into “top-down” \cite{Witten:1998zw,Karch:2002sh,Sakai:2004cn} theories, derived from superstring theory by compactifying extra dimensions, and “bottom-up” \cite{Erlich:2005qh,Gursoy:2007cb,Gursoy:2007er,DeWolfe:2013cua} theories, where a phenomenological gravity model is constructed to reproduce known properties of real matter. In this work, we focus on widely used bottom-up holographic models --- Einstein-Maxwell-Dilaton (EDM) models, which are characterized by the potential of the dilaton field $\cV(\varphi)$ and the coupling between vector and scalar field $\ff_0(\varphi)$.

In bottom-up holographic QCD, two primary approaches are commonly employed:

\begin{enumerate}
    \item The potential reconstruction method \cite{DeWolfe:1999cp,Arefeva:2005mka,He:2010ye,Li:2017tdz,Arefeva:2018hyo, Arefeva:2020byn, Arefeva:2022avn, Arefeva:2022bhx, Chen:2024mmd, Arefeva:2024vom, Arefeva:2024xmg, Arefeva:2024poq, Arefeva:2025xtz, Deng:2026aht} (hereafter called the reconstruction method): One first specifies the geometry (e.g., the warp factor) and the coupling functions $\ff_0(z)$ and then solves the equations of motion, thereby "reconstructing" the dilaton potential $\cV(\varphi)$. In this method the dilaton potential, as a result of the EOM, varies for solutions with  different boundary conditions.

    \item The direct method \cite{DeWolfe:2010he,Knaute:2017opk,Critelli:2017oub,Cai:2022omk,Jokela:2024xgz,Li:2025lmp,Shen:2025yrn,Shen:2025zkj}: One postulates an explicit form for the dilaton potential $\cV(\varphi)$ and the coupling functions $ \ff_0(\varphi)$, and then solves EOM for the metric and field profiles. In this method we find different geometrical factors for different boundary conditions and fixed dilaton potential  $\cV(\varphi)$. 
\end{enumerate}

A natural question arises: which of these methods
is more effective to recover the realistic properties of QCD already observed on the colliders, and produce  the physical predictions –-- such as the phase structure, the equation of state, and the critical behavior.
The investigation of the bottom-up  HQCD, corresponding to QCD  composed of light or heavy quarks, shows a rather effective description within the potential reconstruction method.  
There are  two advantages of the reconstruction method. The light- and heavy-quark 
 models\footnote{In this text, we use the term "light quark model" (or "heavy quark model") to refer to the model obtained by the reconstruction method.} rely on a relatively simple ansatz for the initial functions — namely, the warp factor and the Maxwell gauge-kinetic function, and the equations of motion admit analytical solutions. The same cannot be said of the direct method, where the approximation of the potential and the coupling function can be composed of many parameters and the equations of motion usually admit only numerical solutions.
 \\

While the reconstruction method is highly effective for capturing the phenomeno\-logical properties of QCD under extreme conditions, a subtlety arises in that the reconstructed potential depends on the chosen holographic boundary conditions.
To address this issue, we propose constructing a dilaton potential designed to reproduce the typical features of holographic QCD phase transitions obtained by the reconstruction method, rather than one derived solely from vacuum QCD.  Using a light-quark model as an example, we construct a  potential that captures the principal features of the phase structure of this model.
\\

First, we consider the potential fitted directly from the reconstruction model at values of temperature and chemical potential $(T = 0, \, \mu = 0)$, further labeled as the "vacuum dilaton potential". Since realistic QCD has properties closer to those of the light-quark model, we focused on that case.  This approximation does not capture the phase transitions in a satisfactory way, since this approximation
produces a first-order phase transition (FOPT) at zero chemical potential.
\\

Next, we find a better fit for the potential by minimization of the deviation of the temperature dependence on the horizon position between the direct and reconstruction approaches at zero chemical potential. That deviation is calculated within a finite domain of horizon positions, and a change of that domain allows us to alter the resulting phase diagram and the position of CEP. We present the domain and the corresponding fit, which is selected to reproduce the same position of CEP as in the starting reconstruction model.
\\

The resulting ''minimal dilaton potential'' satisfactorily recovers the physics of the light-quark model even at nonvanishing chemical potential. We call this model 
the minimal-potential model. Specifically, it yields a first-order phase transition (FOPT) line close to those obtained from the reconstruction method. 
The minimal-potential model reproduces a rising Cornell-like quark-antiquark potential up to the hadronic distance $\ell_h=1.5$\,fm, which we use as a phenomenological finite-distance criterion for confinement.
It supports the existence of a phase which can be identified as the quarkyonic phase.
\\

This paper is organized as follows. In Section \ref{s:model}, we present the setup of our holographic model (subsection \ref{ss:act}) and discuss the calculation of thermodynamic variables (subsection \ref{ss:td}) and criteria for phase transitions in the theory (subsection \ref{ss:wl}). In Section \ref{s:rec}, we outline the reconstruction method (subsection \ref{ss:recgen}) and the model under consideration (subsection \ref{ss:lq}). In Section \ref{s:dir}, we describe the method used for calculations in the direct approach, which we label as "shooting". In Section \ref{s:res}, we describe models with the vacuum (subsection \ref{ss:fix}) and the minimal dilaton potential (subsection \ref{ss:min}) and present the resulting phase diagrams of these theories. In Section \ref{s:con}, we review our main results. In Appendix \ref{App:EOM} we list the Equations of Motion for the coordinates used in Section \ref{ss:act}. In Appendix \ref{App:shoot} we expand on the material of Section \ref{s:dir} in more detail, outlining the calculation and the considered Initial Value Problem in more detail (App. \ref{App:IVP}), deriving the expansion of fields near the horizon (App. \ref{App:B}) and asymptotic behavior near the boundary (App. \ref{App:C}). In Appendix \ref{App:D}, we list the potential and coupling functions $\cV_1(\varphi), \ \cV_2(\varphi), \ \ff_0(\varphi) $, used in Section \ref{s:res}.

\section{The holographic setup} \label{s:model}

\subsection{Action of the model} \label{ss:act}

\qquad We consider a 5-dimensional holographic Einstein--Maxwell--Dilaton model \cite{Li:2017tdz} with the Einstein frame action:
\bea \label{action}
S&=&\cfrac{1}{16\pi G_5}\int d^5x\sqrt{- \fg} \left[R-\cfrac{\ff_0(\varphi)}{4}F^2-\cfrac{1}{2}\partial_{\mu}\varphi\partial^{\mu}\varphi-\cV(\varphi)\right],
\eea
where $G_5$ is the 5-dimensional Newtonian  gravitational constant,  $g_{\mu\nu}$ is the Einstein frame metric tensor, $\fg=\mathrm{det}\,g_{\mu\nu}$ is the determinant of the Einstein frame metric tensor, $F_{\mu\nu}$ is the electromagnetic tensor of the gauge Maxwell field $A_{\mu}$, $F_{\mu\nu}=\partial_{\mu}A_{\nu}-\partial_{\nu}A_{\mu}$, $\varphi$ is the dilaton field, $\ff_0(\varphi)$ is the gauge kinetic function associated with the Maxwell field, and $\cV(\varphi)$ is the potential of the dilaton field $\varphi$.

The background is usually described in the Einstein frame($g_{\mu\nu}$), which is connected to the string frame ($g_{\mu\nu}^{(S)}$) via Weyl transformation:
\bea \label{frames}
    g_{\mu\nu}^{(S)} = g_{\mu\nu} e^{\sqrt{\frac{2}{3}}\varphi}.
\eea
The superscript $^{(S)}$ denotes the string frame, whereas its absence corresponds to the Einstein frame.

The general EOMs for the theory \eqref{action} in the Einstein frame take the form:
\bea
    R_{\mu\nu} - \frac{1}{2} R g_{\mu\nu} = T_{\mu\nu},\label{EOMggen}\\
    \Box \varphi - \frac{\partial \cV}{\partial \varphi} - \frac{F^2}{4} \frac{\partial \ff_0}{\partial \varphi} = 0, \label{EOMphigen}\\
    \partial_{\mu} \left(\sqrt{- \fg} \ff_0 F^{\mu\nu} \right) = 0 \label{EOMAgen},
\eea
where $R_{\mu\nu}$ and $R$ are the Ricci tensor and scalar for the Einstein frame metric, $\Box$ is the corresponding Laplace–-Beltrami operator, and $T_{\mu\nu}$ is the energy-momentum tensor for the dilaton and vector fields.

Poincar\'{e} patch coordinates of the deformed AdS space are $x^\mu = \{t,\vec{x}, \zeta \}$, where $t$ is time, $\vec{x}$ is the spatial coordinate, and $\zeta$ is the holographic coordinate. The isotropic ansatz for the metric and fields takes the form:
\bea \label{warp-factor}
ds^2=L^2 e^{2A(\zeta)}\left[-g(\zeta)dt^2+ d\Vec{x}^2\right]+L^2e^{2 B(\zeta)}\cfrac{d\zeta^2}{g(\zeta)},\label{fields_ans}\\
 \quad  \varphi = \varphi(\zeta),\quad A_{\mu} = \left(A_t(\zeta),\Vec{0},0\right),
\eea
where 
$g(\zeta)$ is the blackening function, $A(\zeta)$ and $B(\zeta)$ are the warp factors, and $L$ is the AdS radius. 

For simplicity, we set $L = 1 \ \mathrm{GeV}^{-1}$.

Each field in the ansatz corresponds to a distinct property of the theory:
\begin{itemize}
    \item The blackening function $g(\zeta)$ introduces the temperature;
    \item The electric potential $A_t(\zeta)$ emulates the baryon chemical potential;
    \item The dilaton field $\varphi(\zeta)$ and the warp factors $A(\zeta), \, B(\zeta)$ simulate the running coupling constant of the gauge theory.
\end{itemize}

Owing to the residual covariance under coordinate transformations $\zeta \to \hat{z}(\zeta)$, we can redefine the holographic coordinate. 

We consider two specific types of coordinates: the first is the conformal coordinate $z$, for which $A(z) = B(z)$. In conformal coordinates, the metric \eqref{warp-factor} takes the form
\bea 
ds^2=\cfrac{e^{2 \mcA(z)}}{z^2}\left[-g(z)dt^2+d\Vec{x}^2+\cfrac{dz^2}{g(z)}\right],
\label{metric_z}
\eea
where we also wrote explicitly the unmodified AdS$_5$ scale factor by exchanging $e^{2 A(z)}=\frac{e^{2\mcA(z)}}{z^2}$. The conformal coordinate $z$ is obtained from the general coordinate $\zeta$ via
\bea
    dz^2 = e^{2B(\zeta)-2A(\zeta)}d\zeta^2.
\eea

The second coordinate is the domain wall coordinate $r$, for which $B(r) = 0$. The corresponding metric \eqref{warp-factor} takes the form
\bea \label{dwcoord}
ds^2=e^{2A(r)}\left[-g(r)dt^2+d\Vec{x}^2\right]+\cfrac{dr^2}{g(r)}. \label{mdw}
\eea
The domain wall coordinate $r$ is obtained from the general coordinate $\zeta$ via
\bea
    dr^2 = e^{2B(\zeta)}d\zeta^2.
\eea

\begin{figure}[h!]
  \centering
  \includegraphics[scale=0.8]{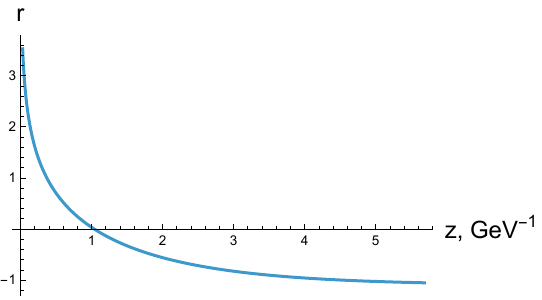}
   \includegraphics[scale=0.8]{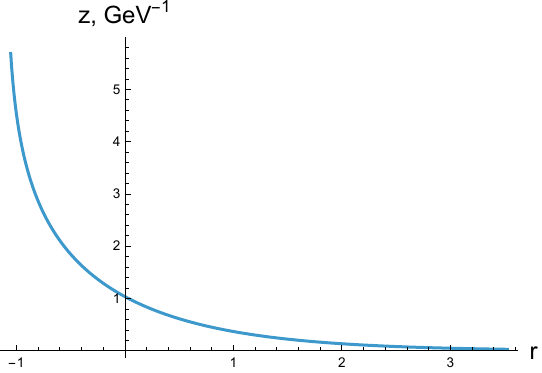}
\caption{Example of the connection between the conformal coordinate $z$ and the domain wall coordinate $r$.}
 \label{Fig:prz}
\end{figure}

The conformal and domain wall coordinates are connected as 
\bea
    dz = -e^{-A(r)}dr.
\eea
Here we explicitly chose the sign $-$, so that as $z$ decreases to $0$, $r$ grows to $\infty$. The interconnection between these coordinates is depicted in Fig. \ref{Fig:prz}.

Equations of motion (EOM) \eqref{EOMggen}-\eqref{EOMAgen} both for the conformal coordinate ansatz \eqref{metric_z} and for the domain wall are listed in App. \ref{App:EOM}.

The conformal (also referred to as UV) boundary of the bulk is located at some value $\zeta_b$, which corresponds to $z \to 0$ or $r \to \infty$. There is also a horizon located at some $\zeta_h$ and corresponding $z_h$ or $r_h$.

We impose boundary conditions (BC). Some of them are:
\bea
    g(\zeta_b) = 1,\text{ --- asymptotically AdS space,}\\
    g(\zeta_h) = 0,\text{ --- the horizon position,}\label{horcon1}\\
    A_t(\zeta_h) = 0,\text{ --- the horizon position,}\label{horcon2}\\
    \varphi(\zeta_b) = 0,\text{ --- imposed by hand to obtain the UV fixed point.}
\eea
The rest of the boundary conditions will be listed in the following sections.

\subsection{Calculation of thermodynamic variables, FOPT}\label{ss:td}

\qquad Thermodynamic properties of the theory are encoded in the solutions and boundary conditions:
\bea
    \mu = A_t(\zeta_b) \text{ --- chemical potential is introduced via vector field \cite{Chamblin:1999tk}}, \label{chempot}\\
   T = \left|\frac{g' e^{A-B}}{4\pi}\right|(\zeta_h) \text{ --- Hawking temperature \cite{Hawking:1974rv}},\label{temp}\\
    S = \frac{e^{3 A(\zeta_h)}}{4 G_5}  \text{ --- Bekenstein-Hawking entropy density \cite{Bekenstein:1973ur, Hawking:1975vcx}.}\label{entropy}
\eea

The considered holographic theory may have a first-order phase transition (FOPT) between black-hole backgrounds of different sizes. This transition starts at some critical endpoint (CEP) and continues along the FOPT line.

To locate the CEP, we consider different states of the background, each parametrized by two parameters $(a_1, \ a_2)$, which vary in different approaches, and examine the map $(a_1, \ a_2) \to (T,\mu)$. CEP can be located as the point at which the map becomes non-invertible (i.e., a state on the phase diagram may be realized in several ways).

The FOPT line position may be calculated in a way similar to the Maxwell construction: calculate the free energy density (up to a $T$-independent constant)
\bea
     F  = -\int \limits_{0}^{T} S(T_1) d T_1, \label{freeen}
\eea
at fixed chemical potential $\mu$, locate and identify points at which $\mu, \ T, \ F$ are equal for two different solutions (large and small black holes).

\subsection{Time-like Wilson loops and the confinement/deconfinement crossover}\label{ss:wl}

\qquad The second considered phase transition is a confinement/deconfinement transition. It is indicated on the gauge side by the expectation value of the time-like Wilson loop \cite{Wilson:1974sk,Maldacena:1998im,Rey:1998ik,Andreev:2006ct} $\langle W(\mathscr{C})\rangle = \langle \mathrm{P}\exp(i g \int_\mathscr{C} dx^\mu A_\mu)\rangle$, oriented along time-like contour $\mathscr{C}$.

The expectation value of the loop is given by the on-shell value of the Nambu-Goto action for the probe string with $\mathscr{C}$ as the world-sheet boundary:
\bea
   \langle W(\mathscr{C})\rangle = e^{-S_{NG}(\mathscr{C})}. 
\eea

The Nambu--Goto action for the string is written in the string frame:
\bea \label{NGa}
    S_{NG} = \frac{1}{2 \pi \alpha'} \int d \xi^1 d \xi^2 \sqrt{-\mathrm{det}h_{ab}},
\eea
where $\xi^a$ are the world-sheet coordinates and the induced metric $h_{ab}, \ a,b = 1,2$ has the form
\bea
    h_{ab} = g^{(S)}_{\mu \nu} \frac{\partial x^\mu}{\partial \xi^a}\frac{\partial x^\nu}{\partial \xi^b},
\eea
with $g^{(S)}_{\mu \nu}$ being the string frame metric components \eqref{frames} and $\alpha'$ being the Regge slope parameter, related to the string tension $
\sigma = 1/2 \pi \alpha'$.

The static quark--antiquark potential is extracted from the renormalized
on-shell Nambu--Goto action according to
\begin{equation}
V_{Q\bar Q}(\ell)=
\lim_{\mathcal T\to\infty}
\frac{S_{\rm NG}^{\rm ren}(\mathcal T,\ell)}{\mathcal T},
\label{eq:QQbar-potential}
\end{equation}
where $\mathcal T$ is the temporal extent of the rectangular
Wilson loop, while $\ell$ is the spatial separation between the quark and
antiquark.
In the following, we omit the overall factor
$1/(2\pi\alpha’)$ and focus on the shape of the renormalized potential.

We choose the parametrization
\bea \label{strpar}
    t \equiv x^0 = \xi^1 \in [0, \ \mathcal T],\\
    x^2 = x^3 = 0, \\
    x^1 \equiv x = \xi^2\equiv \xi \in [-\frac{\ell}{2}, \ \frac{\ell}{2}],\\
    x^4 \equiv \zeta = \zeta(\xi).
\eea

The string frame metric (\ref{frames}) and the Nambu--Goto action (\ref{NGa}) with useful notation take the form 
\bea \left(ds^{(S)}\right)^2=e^{2A^{(S)}(\zeta)}\left[-g(\zeta)dt^2+d\Vec{x}^2\right]+e^{2B^{(S)}(\zeta)}\cfrac{d\zeta^2}{g(\zeta)},\\
\label{NG}
    S_{NG} = \mathcal T\int \limits_{-\frac{\ell}{2}}^{\frac{\ell}{2}} d\xi M(\zeta(\xi)) \sqrt{(\zeta')^2+\mathscr{F}(\zeta(\xi))},\\
    M(\zeta) = \mathrm{exp}(A^{(S)}(\zeta)+B^{(S)}(\zeta)), \label{strnot1}\\
    \mathscr{F}(\zeta) = g(\zeta) \mathrm{exp}(2[A^{(S)}(\zeta)-B^{(S)}(\zeta)]).\label{strnot2}
\eea

We consider a symmetric configuration of a string with a maximum depth into the bulk. This implies the existence of a point $\zeta_0 = \zeta(0)$ (which we call the string bulk depth), at which 
\bea
    \zeta'(\xi)|_{\zeta=\zeta_0} = 0.
\eea

There is an integral of motion for the system \eqref{NG}
\bea
    I = \frac{M(\zeta)}{\sqrt{\mathscr{F}(\zeta)+(\zeta')^2}}\mathscr{F}(\zeta) = M(\zeta_0)\sqrt{\mathscr{F}(\zeta_0)}.
\eea

From it, the distance between quarks can be calculated depending on $\zeta_0$:
\bea \label{sepdist}
    \ell = 2 \int \limits_{\zeta_b}^{\zeta_0} \frac{d\zeta}{\zeta'} =  2 \int \limits_{\zeta_b}^{\zeta_0} \frac{d \zeta }{\sqrt{\mathscr{F}(\zeta)\left(\frac{V^2_{\text{eff}}(\zeta)}{V^2_{\text{eff}}(\tilde{z_0})}-1\right)}}.
\eea
The integration limits $[{\zeta_b},{\zeta_0}]$ are written for the case $\zeta_0 > \zeta_b$, like in the conformal coordinate case \eqref{metric_z}; for the domain wall coordinate \eqref{mdw}, the maximum depth of the string in the bulk is $r_0$ and the integration limits are $[{r_0},{\infty})$.

We introduce the so-called effective potential:
\bea
    V_{\text{eff}}(\zeta) = M(\zeta)\sqrt{\mathscr{F}(\zeta)}. \label{effpot}
\eea

If that potential has extrema, the length of the string can go to infinity, which corresponds to the appearance of the dynamical wall, beyond which the string does not extend; thus, quarks always stay connected --- that is a confined state. If there are no extrema, the string can go up to the horizon and break into two disconnected strings --- that is a deconfined state. Thus, the behavior of the effective potential indicates the crossover phase transition.

The quark-antiquark potential for the string coordinates \eqref{strpar} with notation \eqref{strnot1}-\eqref{strnot2}, \eqref{effpot} can be calculated as
\bea \label{potanz}
V_{Q\bar{Q}} =  2 \int \limits_{\zeta_b}^{\zeta_0}\frac{d \zeta M(\zeta) } {\sqrt{1- \frac{V^2_{\text{eff}}(\zeta_0)}{V^2_{\text{eff}}(\zeta)}}}.
\eea

The integral in \eqref{potanz} diverges at the conformal boundary, since $M(\zeta) \to \infty$ as $\zeta \to \zeta_b$. We renormalize the integral and extract the finite part by subtracting the leading-order divergence independent of the string depth $\zeta_0$; the exact formula for the domain wall coordinate will be written in Section \ref{s:dir}.

\section{The reconstruction method} \label{s:rec}

\subsection{General formulas in the reconstruction method} \label{ss:recgen}
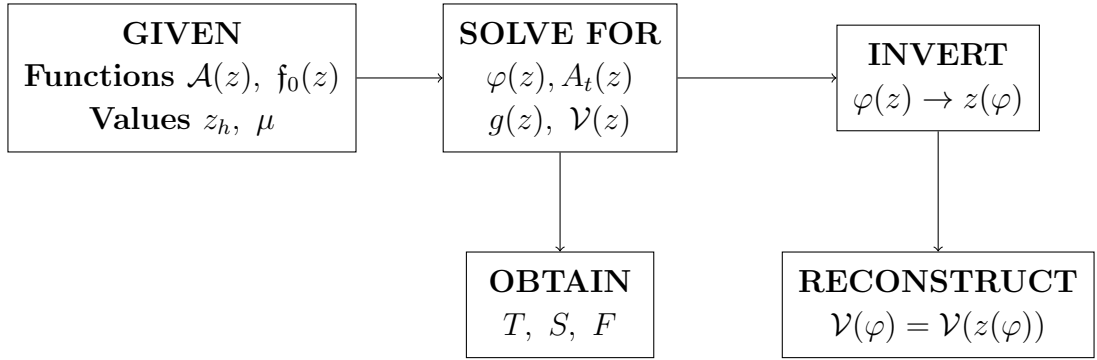
\begin{figure}[h!]
\centering
\begin{tikzpicture}
    \node [block] (Given) {$\begin{array}{ccc}\textbf{GIVEN}\\
    \textbf{Functions } \mathcal{A}(z), \ \ff_0(z)\\
     \textbf{Values } z_h, \ \mu
    \end{array}$};
    \node [block, right of=Given, node distance=5cm] (Solve for) {$\begin{array}{ccc}\textbf{SOLVE FOR}\\
     \varphi(z), A_t(z)\\
      g(z), \ \cV(z)
    \end{array}$};
     \node [block, right of=Solve for, node distance=5cm] (Invert) {$\begin{array}{cc}\textbf{INVERT}\\
     \varphi(z) \to z(\varphi)
    \end{array}$};
    \node [block,below of=Invert, node distance=3cm] (Reconstruct) {$\begin{array}{cc}\textbf{RECONSTRUCT}\\
     \cV(\varphi) = \cV(z(\varphi))
    \end{array}$};
    \node [block, below of=Solve for, node distance=3cm] (Obtain) {$\begin{array}{cc}\textbf{OBTAIN}\\
     T, \ S, \ F
    \end{array}$};
    \draw  [->] (Given) -- (Solve for);
    \draw [->] (Solve for) -- (Invert);
      \draw [->] (Invert) -- (Reconstruct);
    \draw [->] (Solve for) -- (Obtain);
\end{tikzpicture}
\caption{Scheme of the reconstruction method.}\label{fig:recscheme}
\end{figure}

\qquad The reconstruction method uses the conformal coordinate $z \in [0, z_h]$. The routine follows the scheme depicted in Fig. \ref{fig:recscheme}.

Given the function $A(z)$, one can use equation \eqref{phiprime}, together with a boundary condition (e.g., the $\varphi(0) = 0$ condition adopted for the rest of the paper), to solve for the dilaton field:
\bea
    \varphi(z) = \int_0^z d z_1 \sqrt{6\left( -\mathcal{A}''(z_1)+(\mathcal{A}'(z_1))^2-\frac{2}{z_1}\mathcal{A}'(z_1)  \right)} \label{phisol}.
\eea

Given the $\ff_0(z)$ dependence, one can use the chain rule $\ff_{0;\varphi}\varphi' = \ff_0'$ to transform equation \eqref{At2prime} into
\bea
    A''_t+\left(\frac{\ff_0'}{\ff_{0}}+\mathcal{A}'-\frac{1}{z}\right)A'_t=0,
\eea
and solve for the vector field using the  boundary conditions $A_t(0) = \mu, \ A_t(z_h) = 0$:
\bea
A_t(z) = \mu \cfrac{\int_z^{z_h} \frac{y d y}{e^{\mathcal{A}(y)} \ff_0(y)}}{\int_0^{z_h} \frac{x d x}{e^{\mathcal{A}(x)} \ff_0(x)}}.
\eea

With these results, one can use equation \eqref{g2prime}, coupled with boundary conditions $g(0) = 1, \ g(z_h) = 0$, to solve for the blackening function \bea
    g(z) = 1 + \cfrac{1}{\int_0^{z_h}y^3 e^{-3\mathcal{A}(y)}dy} \left( -\int_0^z y^3 e^{-3\mathcal{A}(y)}dy + \left(\cfrac{\mu}{\int_0^{z_h} \frac{x d x}{e^{\mathcal{A}(x)} \ff_0(x)}}\right)^2 G(z,z_h)\right),
\eea
where
\bea
    G(z,z_h) = \int_0^{z_h} y^3 e^{-3 \mathcal{A}(y)} dy \int_{z_h}^{z} x^3 e^{-3 \mathcal{A}(x)} dx \int_0^{x} \frac{t}{e^{\mathcal{A}(t)} \ff_0(t)}dt-\nonumber\\-\int_{z_h}^{z} y^3 e^{-3 \mathcal{A}(y)} dy \int_{0}^{z_h} x^3 e^{-3 \mathcal{A}(x)} dx \int_0^{x} \frac{t}{e^{\mathcal{A}(t)} \ff_0(t)}dt.
\eea

Given all of the fields, it is possible to express the dilaton potential $\cV(z)$ from equation \eqref{A2primes} as
\bea
   \cV(z) = -3 z^2 g e^{-2 \mathcal{A}}\left( \mathcal{A}'' + 3 \mathcal{A}'^2 + 3 \left( \cfrac{3 g'}{2 g} -\cfrac{6}{z}\right)\mathcal{A}'-\cfrac{1}{z}\left(  \cfrac{3 g'}{2 g} -\cfrac{4}{z}\right)+\cfrac{g''}{6g}\right). \label{Vsol}
\eea

Afterwards, one can express the thermodynamic quantities \eqref{temp}-\eqref{freeen} from the fields as
\bea
    T = \cfrac{|g'(z_h)|}{4 \pi},\ S = \cfrac{e^{3 \mathcal{A}(z_h)}}{4 G_5 z_h^3},\\
    F = \int_{z_h}^{z_{h,c}} S(z_{h,1}) T'(z_{h,1}) d z_{h,1},
\eea
where we substituted $d T \to -T' d z_{h,1}$ and introduced the cutoff $z_{h,c}$ corresponding to $T \to 0$.

It is also feasible to numerically invert \eqref{phisol} to obtain the dependence $ z(\varphi)$ and, subsequently,  the potential $\cV(\varphi)$ from \eqref{Vsol}. It should be noted that the resulting potential depends on the considered thermal state, parametrized by $(T,\ \mu)$ or $(z_h,\ \mu)$ values; there is not a single potential in this method.

\subsection{The light quarks model} \label{ss:lq}

\qquad The light quarks model (with modifications) has been extensively studied \cite{Li:2017tdz,Arefeva:2020byn,Arefeva:2022bhx,Arefeva:2022avn,Arefeva:2024vom,Arefeva:2024poq,Arefeva:2024xmg,Arefeva:2025xtz}.

The given functions and parameters are as follows:
\bea
    \mathcal{A}(z) = -a \ln(1+bz^2),\\
    \ff_0(z) = e^{-c z^2 - \mathcal{A}(z)},\\
    G_5 = 1 \ \text{GeV}^{-3}.
\eea

The parameters are chosen to respect the Regge spectrum of the $\rho$--meson \cite{PDBook} ($c = 0.227 \text{ GeV}^2$) and to obtain a confinement/deconfinement transition at $\mu = 0$ close to lattice results \cite{Bazavov:2017dus} ($a = 4.046, \ b = 0.01613 \text{ GeV}^2$).

\begin{figure}[h!]
  \centering
  \includegraphics[scale=0.8]{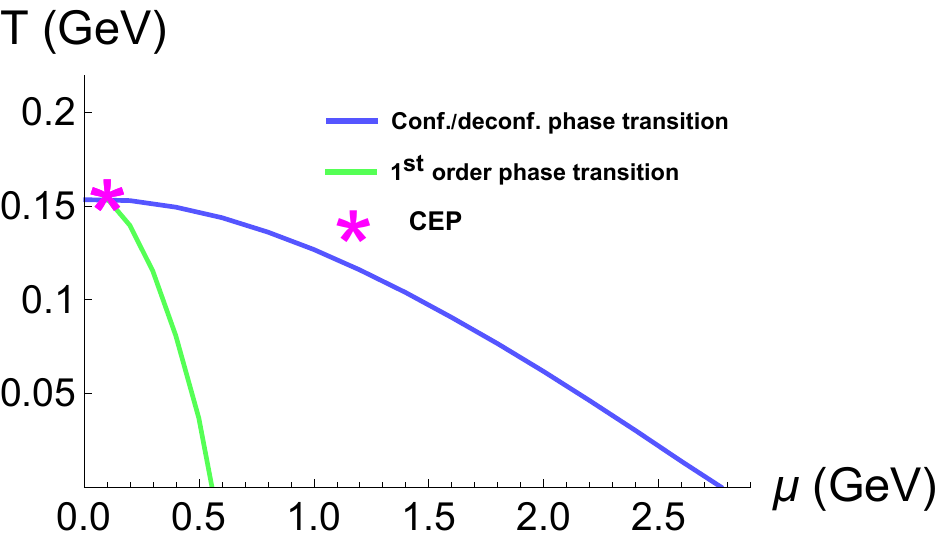}
\caption{Phase diagram in the reconstruction method for the light quarks model, the figure is from the paper \cite{Arefeva:2025xtz}.}
 \label{Fig:recPD}
\end{figure}

The resulting theory has a rich phase structure featuring a FOPT and a confinement/deconfinement crossover (see Fig. \ref{Fig:recPD}). The FOPT line ends at the CEP $(T, \ \mu) = ( 0.1578, \ 0.04779) \text{ GeV}$, while the two phase transition lines intersect at $(T, \ \mu) = (0.1538, \ 0.1043) \text{ GeV}$.

\clearpage

\section{Direct method } \label{s:dir}

\begin{figure}[h!]
\centering
\begin{tikzpicture}
    \node [block] (Given) {$\begin{array}{ccc}\textbf{GIVEN}\\
    \textbf{Functions } \cV(\varphi), \ \ff_0(\varphi)\\
     \textbf{Values } \varphi_h, \ a_1
    \end{array}$};
    \node [block, right of=Given, node distance=5cm] (Solve for) {$\begin{array}{ccc}\textbf{SOLVE FOR}\\
     \varphi(r), A_t(r)\\
      g(r), \ A(r)
    \end{array}$};
    \node [block, right of=Solve for, node distance=4cm] (Obtain) {$\begin{array}{cc}\textbf{OBTAIN}\\
     T, \ S, \ F
    \end{array}$};
    \draw  [->] (Given) -- (Solve for);
    \draw [->] (Solve for) -- (Obtain);
    
\end{tikzpicture}
\caption{Scheme of the shooting calculation in the direct method.} \label{fig:shscheme}
\end{figure}
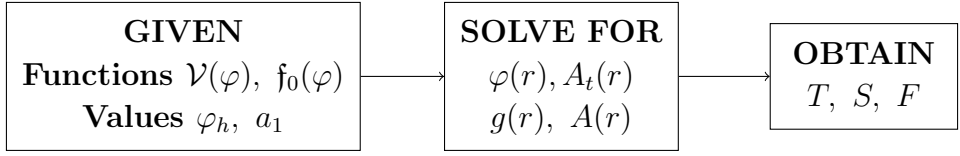

\qquad The shooting calculation \cite{DeWolfe:2010he, Knaute:2017opk, Critelli:2017oub,Jokela:2024xgz} works with the domain wall coordinate $r \in [r_h, \ \infty)$ (or some analog), rather than with the conformal coordinate $z$. It follows the scheme (first introduced in \cite{DeWolfe:2010he}) depicted in Fig. \ref{fig:shscheme}.

We intend to start at the horizon and numerically solve the EOM to get to the conformal boundary of the theory, thus "shooting" from one boundary to another.
\begin{figure}[h!]
  \centering
  \includegraphics[scale=0.7]{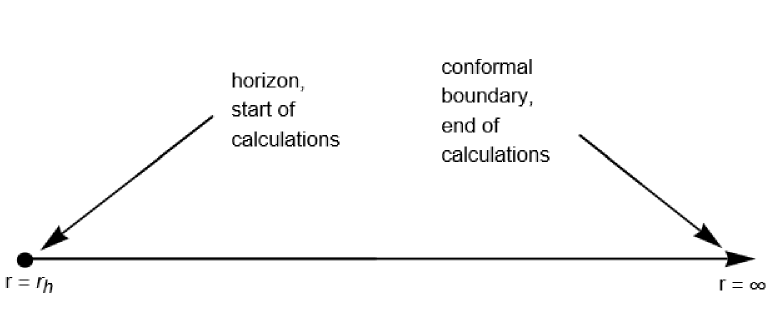} 
\caption{Coordinate axis showing the start and end point of the calculations.}
 \label{Fig:r-axis0}
\end{figure}

A more detailed explanation of the shooting calculation is provided in App. \ref{App:shoot}.

The most important parameters used in the shooting calculations are encoded in the UV asymptotics of the theory, in the region where $\varphi \to 0$. They are determined by the leading order dilaton potential $\cV(\varphi)$ expansion and the $\varphi(r)$ behavior:
\bea
    \cV(\varphi) \simeq -12 + \cfrac{1}{2}m^2 \varphi^2,\\
    \varphi(r) = \varphi_A e^{ - \upnu r}.
\eea
The parameters are the dilaton mass $m^2$ and the corresponding conformal dimension of the dilaton $\upnu = \Delta_- =2 - \sqrt{4 + m^2}$, as well as a characteristic energy scale of the theory $ \Lambda_{es} = \varphi_A^{\frac{1}{\upnu}}$, which is a proxy for the AdS radius $L = 1$ GeV$^{-1}$ fixed earlier.

The formula for the quark-antiquark potential \eqref{potanz} in the domain wall coordinate \eqref{dwcoord} takes the form
\bea
    V_{Q\bar{Q}} =  2 \int \limits_{r_0}^{\infty}\frac{d r M(r) } {\sqrt{1- \frac{V^2_{\text{eff}}(r_0)}{V^2_{\text{eff}}(r)}}} = 2 \int \limits_{r_0}^{\infty}\frac{d r e^{A(r) +  \sqrt{\frac{2}{3}} \varphi(r)} } {\sqrt{1- \frac{g(r_0)e^{4 A(r_0) + 2 \sqrt{\frac{2}{3}}\varphi(r_0)}}{g(r)e^{4 A(r) + 2 \sqrt{\frac{2}{3}}\varphi(r)}}}},
\eea
with $r_0$ being the depth of string into the bulk. Taking into account the asymptotics from App. \ref{App:shoot}, we get that the integral diverges on the upper limit as
\bea \label{divas}
V_{Q\bar{Q}} \sim \lim_{r \to \infty}2 \int \limits^r_{r_0} e^r = \lim_{r \to \infty} \left(2 e^r - 2 e^{r_0}\right).
\eea

We obtain formula for the renormalized potential by subtracting the first term from \eqref{divas}:
\bea
    V_{Q \bar{Q}}^{{R}} =2 \int \limits_{r_0}^{\infty}dr\left(\frac{ e^{A(r) + \sqrt{\frac{2}{3}} \varphi(r)} } {\sqrt{1- \frac{g(r_0) e^{4 A(r_0) + 2\sqrt{\frac{2}{3}} \varphi(r_0)}}{g(r) e^{4 A(r) + 2\sqrt{\frac{2}{3}} \varphi(r)}}}} - e^r\right)-2e^{r_0}.
\eea

Formula for the quark-antiquark separation \eqref{sepdist} distance takes the form:
\bea
    \ell = 2 \int \limits_{r_0}^{\infty} \frac{d r }{\sqrt{1- \frac{g(r_0) e^{4 A(r_0) + 2\sqrt{\frac{2}{3}} \varphi(r_0)}}{g(r) e^{4 A(r) + 2\sqrt{\frac{2}{3}} \varphi(r)}}}}.
\eea

\section{Results for the direct method based on the reconstruction model}\label{s:res}

\qquad In this section, we aim to reproduce the results from the previously discussed reconstruction method light quarks model (subsection \ref{ss:lq}) using the direct method. 

As per the methods discussed in Section \ref{s:dir}, some a priori parameters need to be provided. These are:
\bea
    \upnu = 1, \ \Delta = 3,\\
    \Lambda_{es} = 1.533 \text{ GeV}.
\eea

We also need to fix the dilaton potential and coupling functions $\cV(\varphi)$ and $\ff_0(\varphi)$. Used fits are described in this section and listed in Appendix \ref{App:D}. 

The coupling function $\ff_0(\varphi)$ is obtained by using the $\ff_0(z)$ and the $\varphi(z)$ dependencies from subsection \ref{ss:recgen} for the model in subsection \ref{ss:lq}. Both of these functions do not depend on $z_h$ or $\mu$, so the coupling function remains fixed for all solutions considered in the reconstruction method.

The resulting $\cV(\varphi)$ function depends on $g(z)$ and thus is implicitly affected by the values of $z_h, \ \mu$; as it has been noted in subsection \ref{ss:recgen}, there is not a single potential. Thus, different fitting methods and functions $\cV(\varphi)$ may lead to differing results and models. In the following subsections \ref{ss:fix}, \ref{ss:min}, we describe two methods we employed for obtaining the dilaton potential. In subsection \ref{ss:phasedg} we present the resulting phase diagram.

We seek it in the form similar to one commonly used by other authors \cite{DeWolfe:2010he,Critelli:2017oub,Jokela:2024xgz}:
\bea \label{potform}
     \cV(\varphi,\{a_n\}) = -12 \cosh(a_1 \varphi) + (6 a_1^2 - 1.5) \varphi^2 + \sum_{i=2}^{9} a_i \varphi^{2i},
\eea
where $\{a_n\}$ is a list of parameters to be calculated. The $ (6 a_1^2 - 1.5) \varphi^2$ term is introduced to reproduce the UV asymptotics of the dilaton \eqref{UV-V}. We introduced more terms in the polynomial expansion in \eqref{potform}  than used in works of other authors to facilitate accuracy of fitting the starting reconstruction potential. Also we note that significantly reducing the number of terms (so that order of the last one is $\varphi^{14}$ or lower) prohibited us from reproducing the non-vanishing CEP position in subsection \ref{ss:min}.

We would like to list parameters describing the numerical procedure for both methods from Section \ref{s:dir}. Throughout the shooting calculations, we started our IVP solution at $\epsilon = 10^{-14}$, used the series expansion from Appendix \ref{App:B} up to $N = 5$ and stopped the IVP integration after values of the dilaton $\varphi_1 = 0.1$ and $\varphi_2 = 0.05$. All resulting differential equations were solved on Wolfram Mathematica.

\subsection{Results for the vacuum dilaton potential} \label{ss:fix} 

\qquad We fixed the parameters $\{a_n\}$ from \eqref{potform} in two ways. First, we calculated the reconstruction potential $\cV(\varphi)$ corresponding to some values of $(T, \ \mu)$ (that is, $(z_h, \ \mu)$) in the light quarks model (subsection \ref{ss:lq}) and obtained a fit for the parameters.

We fitted the parameters for the potential calculated for the vacuum thermal state at $(T, \ \mu) = (0 \text{ GeV}, 0 \text{ GeV})$. Finite $(T, \ \mu)$ change values slightly, but the results remain quantitatively similar. The corresponding fit for the potential is labeled as $\cV_1(\varphi)$ and is presented in Appendix \ref{App:D}. We refer to that potential as the ''vacuum dilaton potential''.

\begin{figure}[h!]
  \centering

 \begin{subfigure}{0.45 \textwidth}
     \includegraphics[scale=0.725]{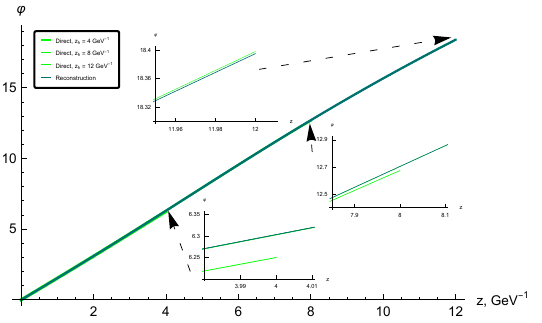}
 \end{subfigure}
\hspace{2em}
\begin{subfigure}{0.45 \textwidth}
    \includegraphics[scale=0.725]{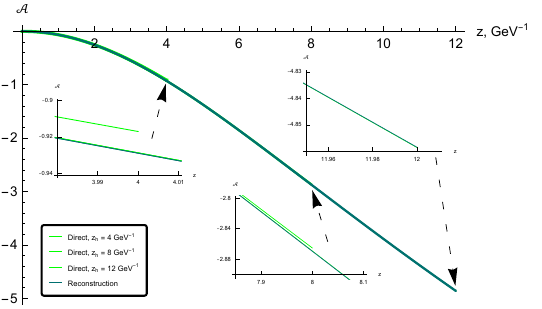}
\end{subfigure}
 \begin{subfigure}{0.45 \textwidth}
     \includegraphics[scale=0.725]{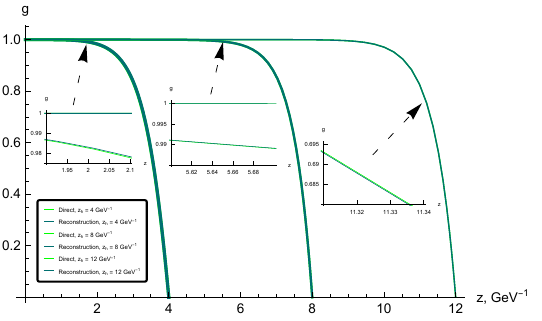}
 \end{subfigure}
\hspace{3em}
\begin{subfigure}{0.45 \textwidth}
    \includegraphics[scale=0.725]{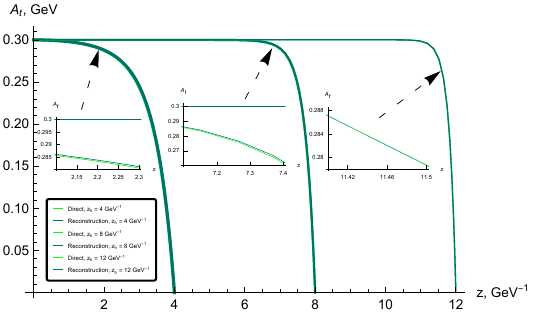}
\end{subfigure}

\caption{Solutions for the dilaton field, the warp factor, the blackening function and the vector potential for the reconstruction method and the direct method with vacuum dilaton potential. Chemical potential is fixed at $0.3$ GeV, while the horizon positions are $z_h \in \{ 4, \, 8, \, 12\}$ GeV$^{-1}$. Some regions are enlarged in the insets. Throughout, $z$ is measured in GeV$^{-1}$, $A_t$ in GeV and other quantities are dimensionless.}
 \label{Fig:solold}
\end{figure}

Comparison of solutions for the light quarks model and the direct method with vacuum potential is depicted in Fig. \ref{Fig:solold} for values of chemical potential $\mu = 0.3$ GeV and positions of the horizon $z_h = 4$ GeV$^{-1}$, $8$ GeV$^{-1}$, $12$ GeV$^{-1}$. We note that the warp-function $A(z)$ and the dilaton field $\varphi(z)$ do not depend on the chemical potential or the horizon position for the reconstruction result.

The solutions grow closer as $z_h$ grows, as can be inferred from the plots. The difference is the greatest for values of $z_h$ near the location of spurious FOPT, which is described further.

\begin{figure}[h!]
\centering
\includegraphics[scale=1.1]{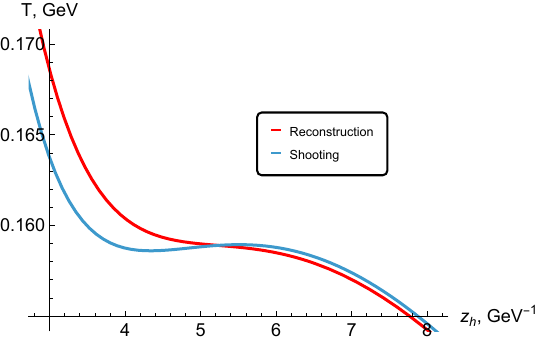} 
\caption{$T(z_h) $ dependence for the chemical potential $\mu = 0 \text{ GeV}$ for the reconstruction method and the direct method with the vacuum dilaton potential. Note the FOPT.} \label{foptold}
\end{figure}

The phase structure of the resulting theory differed from the reconstruction case: there was a FOPT at $\mu = 0 \text{ GeV}$ (depicted in Fig. \ref{foptold}), which was absent before (see subsection \ref{ss:lq}). At the same time, predictions of the reconstruction and direct model converge at greater $\varphi_h$.

\begin{figure}[h!]
\includegraphics[scale=1.4]{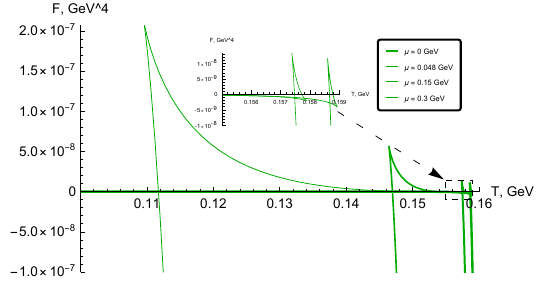} 
\caption{Plots of free energy density $F$ versus temperature $T$ for different values of chemical potential $\mu = 0 \text{ GeV}, \, 0.048\text{ GeV}, \, 0.15\text{ GeV}, \, 0.3$ GeV for the vacuum dilaton potential.}

 \label{Fig:FTold}
\end{figure}

The corresponding dependence of the free energy over temperature $F(T)$ is depicted in Fig. \ref{Fig:FTold}. We see that $F(T)$ isn't single-valued for all chemical potentials $\mu$ and the plots exhibit a swallow-tail behavior, which grows larger and is shifted to smaller temperatures as the chemical potential grows.

\begin{figure}[h!]
\includegraphics[scale=1.1]{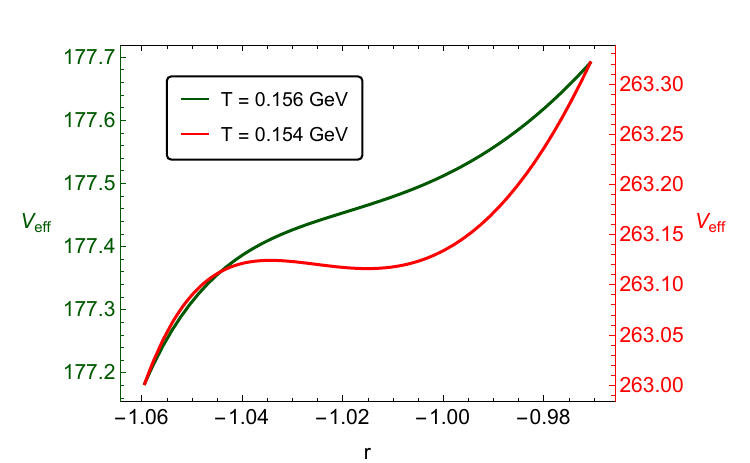} 
\caption{Effective potential in confined ($T = 0.154$ GeV) and deconfined ($T = 0.156$ GeV) phases for chemical potential $\mu = 0$ GeV for the vacuum dilaton potential.
}
 \label{Fig:EffPotold}
\end{figure}

\begin{figure}[h!]
  \centering
 $\mu=0$ GeV\hspace{150pt} $\mu=0.048$ GeV\\
 \begin{subfigure}{0.45 \textwidth}
     \includegraphics[scale=0.9]{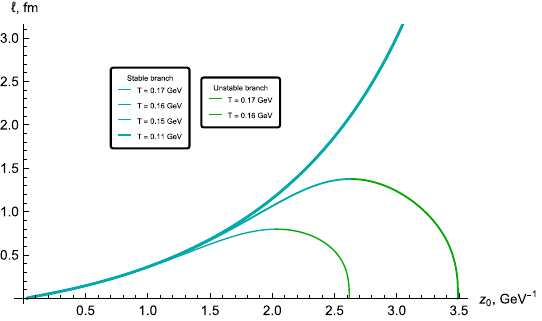}
     \subcaption{A}
 \end{subfigure}
\hfill
\begin{subfigure}{0.45 \textwidth}
    \includegraphics[scale=0.9]{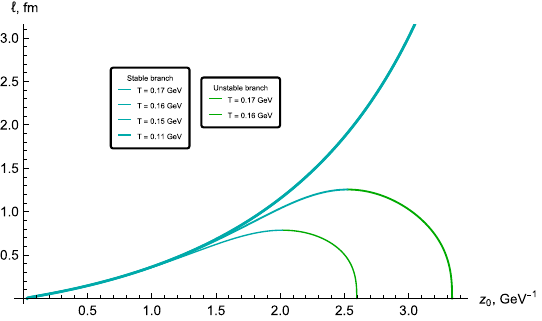}
    \subcaption{B}
\end{subfigure}\\
\caption{Distance between quarks as a function of depth of string in the bulk for the vacuum dilaton potential at temperatures $T = 0.17\text{ GeV}, \, 0.16\text{ GeV}, \, 0.15\text{ GeV}, \, 0.11$ GeV for fixed $\mu = 0 \text{ GeV (A)}, \, \mu = 0.048 \text{ GeV (B)} $. Temperatures $T = 0.17\text{ GeV}, \, 0.16$ GeV correspond to the deconfined phase, while temperatures $T = 0.15\text{ GeV}, \, 0.11$ GeV correspond to the confined phase.}
 \label{Fig:QQLzoldfixmu}
\end{figure}

\begin{figure}[h!]
  \centering
 $\mu=0$ GeV\hspace{150pt} $\mu=0.048$ GeV\\
 \begin{subfigure}{0.45 \textwidth}
     \includegraphics[scale=0.9]{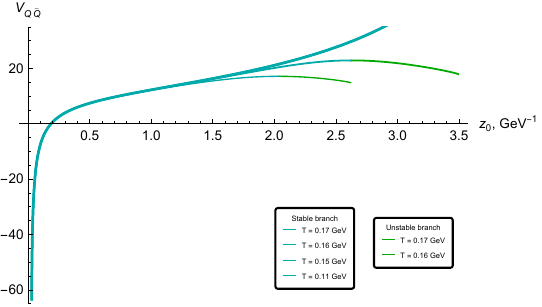}
     \subcaption{A}
 \end{subfigure}
\hfill
\begin{subfigure}{0.45 \textwidth}
    \includegraphics[scale=0.9]{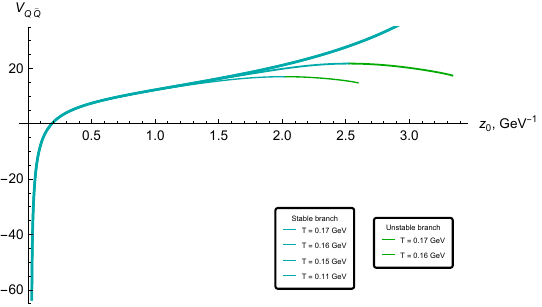}
    \subcaption{B}
\end{subfigure}\\
\caption{Quark-antiquark potential as a function of depth of string in the bulk for the vacuum dilaton potential at temperatures $T = 0.17\text{ GeV}, \, 0.16 \text{ GeV}, \, 0.15\text{ GeV}, \, 0.11$ GeV for fixed $\mu = 0 \text{ GeV (A)}, \, \mu = 0.048 \text{ GeV (B)} $. Temperatures $T = 0.17\text{ GeV}, \, 0.16$ GeV correspond to the deconfined phase, while temperatures $T = 0.15\text{ GeV}, \, 0.11$ GeV correspond to the confined phase.}
 \label{Fig:QQVzoldfixmu}
\end{figure}

\begin{figure}[h!]
  \centering
 $\mu=0$ GeV\hspace{150pt} $\mu=0.048$ GeV\\
 \begin{subfigure}{0.45 \textwidth}
     \includegraphics[scale=0.9]{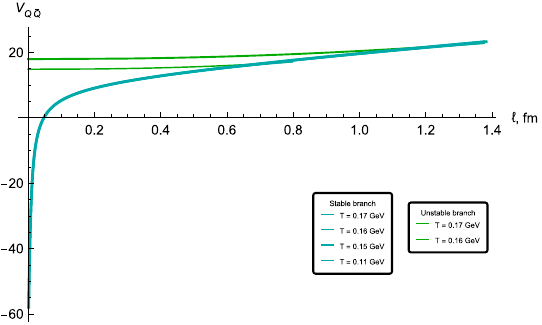}
     \subcaption{A}
 \end{subfigure}
\hfill
\begin{subfigure}{0.45 \textwidth}
    \includegraphics[scale=0.9]{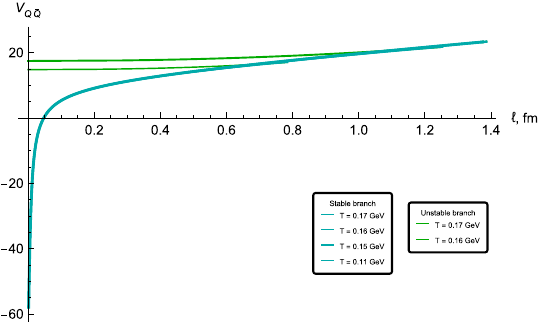}
    \subcaption{B}
\end{subfigure}\\
\caption{Quark-antiquark potential for the vacuum dilaton potential at temperatures $T = 0.17 \text{ GeV}, \, 0.16\text{ GeV}, \, 0.15\text{ GeV}, \, 0.11$ GeV for fixed $\mu = 0 \text{ GeV (A)}, \, \mu = 0.048 \text{ GeV (B)} $. Temperatures $T = 0.17 \text{ GeV}, \, 0.16$ GeV correspond to the deconfined phase, while temperatures $T = 0.15 \text{ GeV}, \, 0.11$ GeV correspond to the confined phase.}
 \label{Fig:QQoldfixmu}
\end{figure}

There is a confinement/deconfinement crossover, which can be seen in the effective potential plots in Fig. \ref{Fig:EffPotold}. The dependencies of the distance between quark and antiquark $\ell$ and the quark-antiquark potential $V_{Q\bar{Q}}$ on the string bulk depth $z_0$ for the direct method with the vacuum dilaton potential are depicted in Fig. \ref{Fig:QQLzoldfixmu} and \ref{Fig:QQVzoldfixmu}.
The corresponding dependence of quark-antiquark potential $V_{Q\bar{Q}}$ on the distance between quark and antiquark $\ell$ in different phases is depicted in Fig. \ref{Fig:QQoldfixmu}.

For the confined phase ($T = 0.15 \text{ GeV}, \, 0.11$ GeV on the figure), the $\ell(z_0)$ and $V_{Q\bar{Q}}(z_0)$ dependencies are monotonous functions increasing to infinity, while the potential $V_{Q\bar{Q}}(\ell)$ is a single-valued function exhibiting a linear growth for $\ell \to \infty$. For the deconfined phase ($T = 0.17 \text{ GeV}, \, 0.16$ GeV on the figure), the $\ell(z_0)$ and $V_{Q\bar{Q}}(z_0)$ dependencies have a maximum point before the horizon and decline after reaching it, while the potential $V_{Q\bar{Q}}(\ell)$ is a double-valued function. We label the branch with the lesser potential $V_{Q\bar{Q}}$ at fixed $\ell$ as a "stable branch", while the other is labeled as an "unstable branch". The stable branch of the potential exhibits a linear growth at large $\ell$, but it bounces back via the unstable branch after reaching a maximum point. Different values of chemical potential $\mu$ or temperature $T$ affect the behavior of functions in a confined phase only slightly, while in a deconfined phase different thermal backgrounds correspond to different maximum points and lower $\ell$ and $V_{Q\bar{Q}}$ values overall.

\clearpage

\subsection{Results for the minimal dilaton potential} \label{ss:min}

\qquad As we aim to reproduce the phase structure following the reconstruction method, we need to obtain a similar CEP at $\mu \ne 0 \text{ GeV}$. To that end, we searched for the potential parameters $\{ a_n \}$ as a result of minimization of a functional, which defines the difference with the reconstruction results. We minimized the integrated $T(z_h)$ deviation between the reconstruction method (labeled as $T_{rec}(z_h)$) and the direct method (labeled as $T_{dir}(z_h,\{ a_n \})$ and depending on parameters $\{ a_n \}$)  in some domain $[z_{h1}, \, z_{h2}]$:
\bea
    \Delta T_{int}(\{ a_n \}) = \int \limits_{z_{h1}}^{z_ {h2}} dz_h \ (T_{rec}(z_h)-T_{dir}(z_h,\{ a_n \}))^2.
\eea

Minimization of that functional in different domains resulted in theories with different CEPs.  Positions of CEP values corresponding to several fit domains are depicted in Table \ref{t:CEPS}.

\begin{table}[h!]
\begin{center}
	\begin{tabular}{|c|c|c|c|c|c|}
		\hline
		&$z_h \in [2, \, 8.25]$ & $z_h \in [2, \, 8.375]$&$z_h \in [2, \, 8.5]$&$z_h \in [2, \, 9]$ &$z_h \in [2, \, 10]$   \\
		\hline
		 CEP $(T, \, \mu)$& $(0.1577,0.053)$& $(0.158, \, 0.048)$&$(0.1583,0.041)$& $(\dots,\, 0)$&$(\dots,\, 0)$\\
		\hline
	\end{tabular}
\end{center}
\caption{CEP positions corresponding to different fitting domains. For fits with a finite CEP, its temperature and chemical potential are presented as $(T, \, \mu)$, while for the fits without CEP (or with CEP at vanishing chemical potential) the result is presented as $(\dots,\, 0)$. Temperature and chemical potential $(T, \, \mu)$ are measured in GeV, while horizon position $z_h$ is measured in GeV$^{-1}$.}
\label{t:CEPS}
\end{table}

As can be inferred from the table, the position of the CEP varies greatly depending on the fitting domain. It depends on the IR region and the degree of correspondence to the reconstruction results which we impose: if the fitting domain extends too deep into the IR region with $z_h \to \infty$, the fit results in correspondence to the vacuum potential case. In that case, the spurious FOPT appears, as can be seen from the fits with $z_h \ge 9$ GeV$^{-1}$. 

Our best result was for the domain $z_h \in [2, \, 8.375]$ GeV$^{-1}$, for which we obtained a potential fit $\cV_2(\varphi)$, which is presented in Appendix \ref{App:D}. This fit had the CEP position (specifically, the chemical potential) closest to the starting reconstruction case CEP. We refer to that potential as the ''minimal dilaton potential'', and we refer to the corresponding HQCD model as the ''minimal--potential model''.

\begin{figure}[h!]
  \centering

 \begin{subfigure}{0.45 \textwidth}
     \includegraphics[scale=0.725]{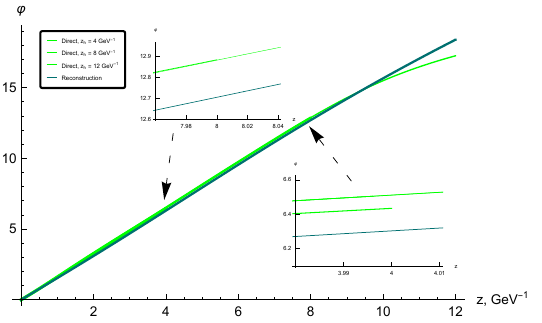}
 \end{subfigure}
\hspace{2em}
\begin{subfigure}{0.45 \textwidth}
    \includegraphics[scale=0.725]{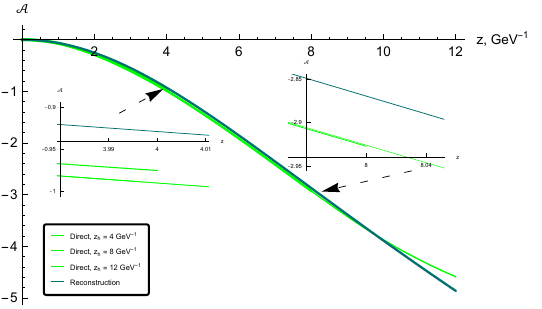}
\end{subfigure}
 \begin{subfigure}{0.45 \textwidth}
     \includegraphics[scale=0.725]{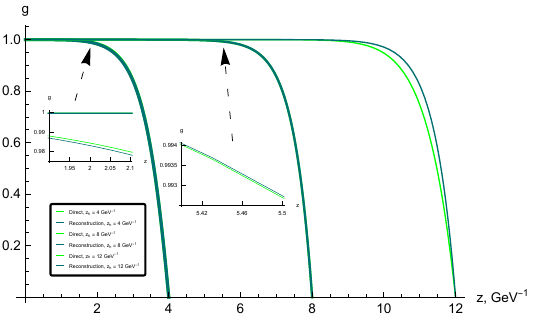}
 \end{subfigure}
\hspace{3em}
\begin{subfigure}{0.45 \textwidth}
    \includegraphics[scale=0.725]{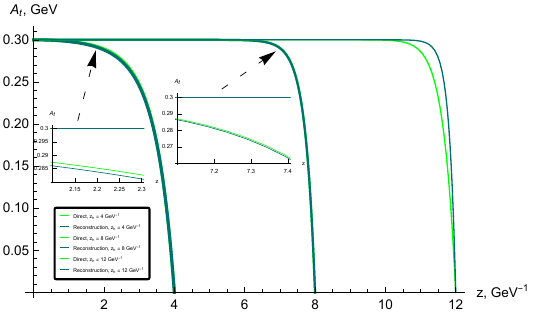}
\end{subfigure}

\caption{Solutions for the dilaton field, the warp factor, the blackening function and the vector potential for the reconstruction method and the direct method with minimal potential. Chemical potential is fixed at $0.3$ GeV, while the horizon positions are $z_h \in \{ 4, \, 8, \, 12\}$ GeV$^{-1}$. Some regions are enlarged in the insets. Throughout, $z$ is measured in GeV$^{-1}$, $A_t$ in GeV and other quantities are dimensionless.}
 \label{Fig:solmin}
\end{figure}

Comparison of solutions for the light quarks model and the direct method with the minimal dilaton potential is depicted in Fig. \ref{Fig:solmin} for values of chemical potential $\mu = 0.3$ GeV and positions of the horizon $z_h = 4$ GeV$^{-1}$, $8$ GeV$^{-1}$, $12$ GeV$^{-1}$. We note that the warp-function $A(z)$ and the dilaton field $\varphi(z)$ do not depend on the chemical potential or the horizon position for the reconstruction result.

Solutions from different methods are the closest for values of $z_h$ in the fitting domain. As can be inferred from the plot of dilaton field or warp factor, the solutions for different methods diverge as $z_h$ grows.

\begin{figure}[h!]
\centering
\includegraphics[scale=1.5]{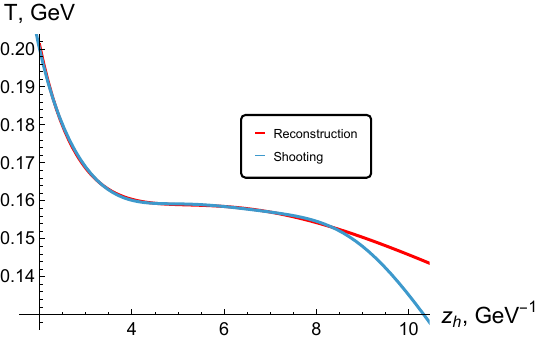} 
\caption{$T(z_h) $ dependence for the chemical potential $\mu = 0 \text{ GeV}$ for the reconstruction method and the direct method with the minimal dilaton potential. Note the absence of FOPT.} \label{Tznew}
\end{figure}

\begin{figure}[h!]
\includegraphics[scale=1.2]{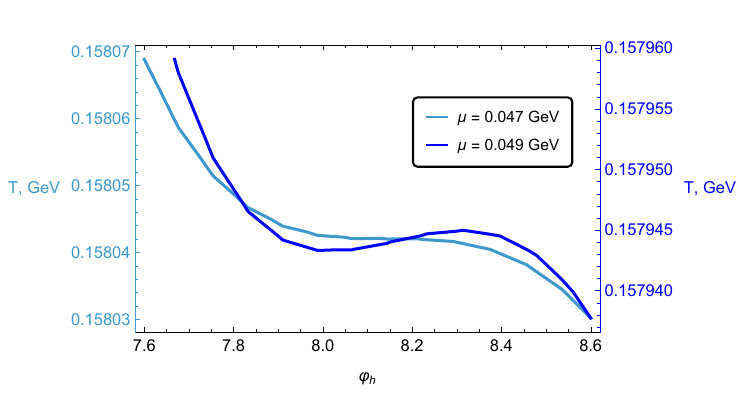} 
\caption{$T(\varphi_h)$ dependence for the minimal dilaton potential at $\mu = 0.047$ GeV and $\mu = 0.049$ GeV --- right before and after the CEP.}

 \label{Fig:CEPnew}
\end{figure}

The $T(z_h)$ dependence in the corresponding theory follows the reconstruction results closely until $z_h = 8.375$ GeV$^{-1}$, after which it differs greatly, as can be seen in Fig. \ref{Tznew}. The corresponding theory has CEP values of $(T, \ \mu) = (0.158  \ \text{GeV}, \ 0.048 \ \text{GeV})$ --- close to the reconstruction result  $(T, \ \mu) = (0.1578  \ \text{GeV}, \ 0.04779 \ \text{GeV})$, see Fig. \ref{Fig:CEPnew}. 

\begin{figure}[h!]
\includegraphics[scale=1.4]{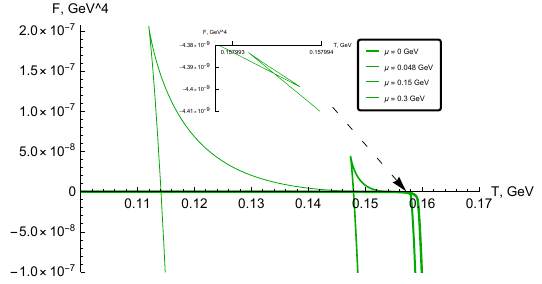} 
\caption{Plots of free energy density $F$ versus temperature $T$ for different values of chemical potential $\mu = 0 \text{ GeV}, \, 0.048 \text{ GeV}, \, 0.15 \text{ GeV}, \, 0.3$ GeV for the minimal dilaton potential.}

 \label{Fig:FT}
\end{figure}

The corresponding dependence of the free energy versus temperature $F(T)$ is depicted in Fig. \ref{Fig:FT}. We see that $F(T)$ is single-valued for small $\mu$ values before the CEP while for greater $\mu$
plots exhibit a swallow-tail behavior, which grows larger and is shifted to smaller
temperatures as the chemical potential grows.

\begin{figure}[h!]
\includegraphics[scale=1.3]{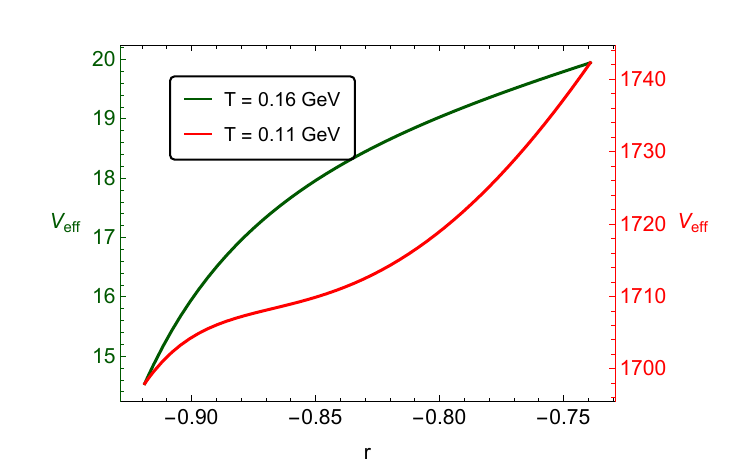} 
\caption{Effective potential for temperatures ($T = 0.16 \text{ GeV}, \, 0.11$ GeV) and chemical potential $\mu = 0 \text{ GeV}$ for the minimal dilaton potential.}
 \label{Fig:EffPotnew}
\end{figure}

 There was no dynamical wall in the theory. The effective potential exhibited monotonic behavior (see Fig. \ref{Fig:EffPotnew}). Therefore, for this model we
adopt a phenomenological finite-distance criterion. We regard a quark--antiquark pair as confined when the rising Cornell-like behavior of $V_{Q\bar Q}(\ell)$ persists up to the hadronic distance $\ell=1.5\,\mathrm{fm}$. This criterion characterizes the potential on hadronic length scales; it is not a statement about an asymptotic area law at $\ell\to\infty$.

\begin{figure}[h!]
  \centering
 $\mu=0$ GeV\hspace{150pt} $\mu=0.048$ GeV\\
 \begin{subfigure}{0.45 \textwidth}
     \includegraphics[scale=0.9]{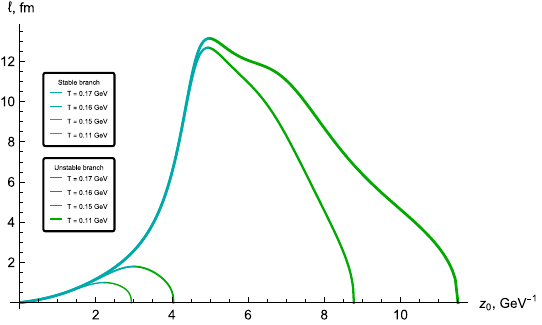}
     \subcaption{A}
 \end{subfigure}
\hfill
\begin{subfigure}{0.45 \textwidth}
    \includegraphics[scale=0.9]{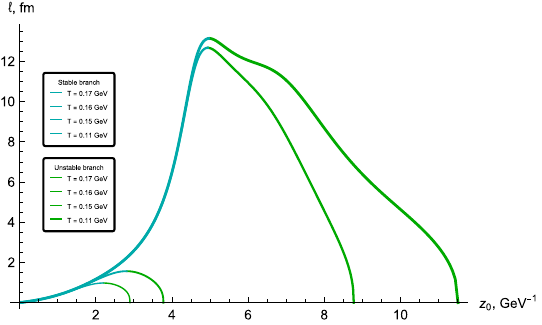}
    \subcaption{B}
\end{subfigure}\\\,\\
 $\mu=0.15$ GeV\hspace{150pt} $\mu=0.3$ GeV\\
 \begin{subfigure}{0.45 \textwidth}
     \includegraphics[scale=0.9]{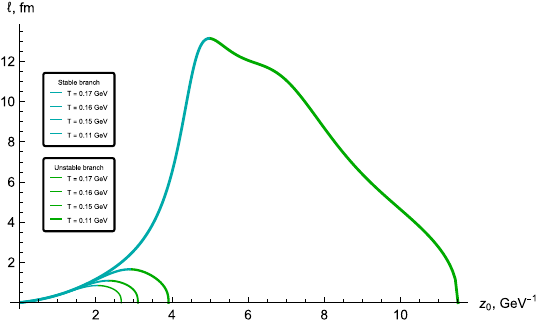}
     \subcaption{C}
 \end{subfigure}
\hfill
\begin{subfigure}{0.45 \textwidth}
    \includegraphics[scale=0.9]{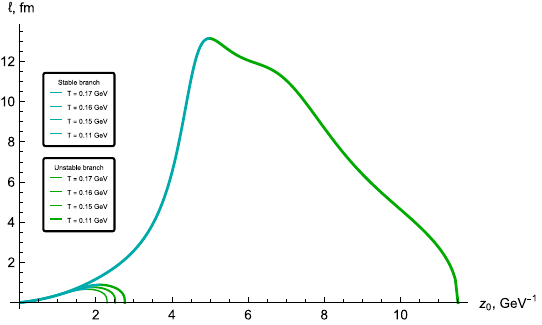}
    \subcaption{D}
\end{subfigure}

\caption{Distance between quarks as a function of depth of string in the bulk for the minimal dilaton potential at temperatures $T = 0.17 \text{ GeV}, \, 0.16 \text{ GeV}, \, 0.15 \text{ GeV}, \, 0.11 \text{ GeV}$ for fixed $\mu = 0 \text{ GeV (A)}, \, \mu = 0.048 \text{ GeV (B)}, \,\mu = 0.15 \text{ GeV (C)}, \,\mu = 0.3 \text{ GeV (D)}$.}
 \label{Fig:QQLzfixmu}
\end{figure}

\begin{figure}[h!]
  \centering
 $\mu=0$ GeV\hspace{150pt} $\mu=0.048$ GeV\\
 \begin{subfigure}{0.45 \textwidth}
     \includegraphics[scale=0.9]{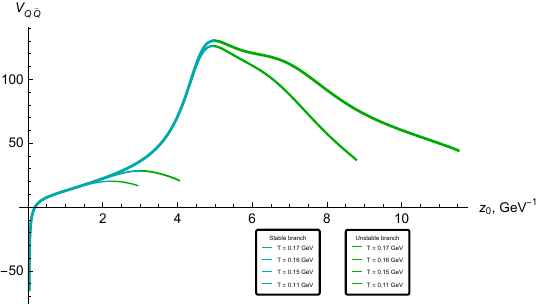}
     \subcaption{A}
 \end{subfigure}
\hfill
\begin{subfigure}{0.45 \textwidth}
    \includegraphics[scale=0.9]{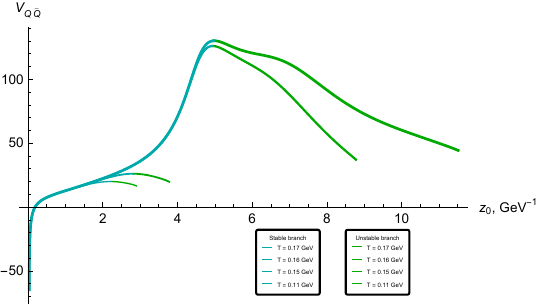}
    \subcaption{B}
\end{subfigure}\\\,\\
 $\mu=0.15$ GeV\hspace{150pt} $\mu=0.3$ GeV\\
 \begin{subfigure}{0.45 \textwidth}
     \includegraphics[scale=0.9]{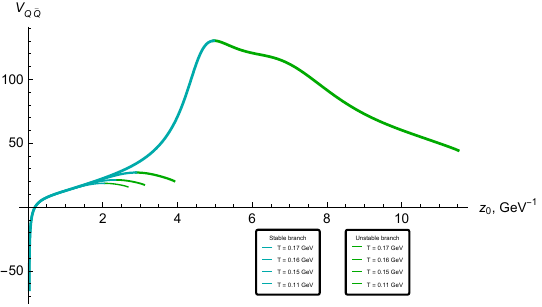}
     \subcaption{C}
 \end{subfigure}
\hfill
\begin{subfigure}{0.45 \textwidth}
    \includegraphics[scale=0.9]{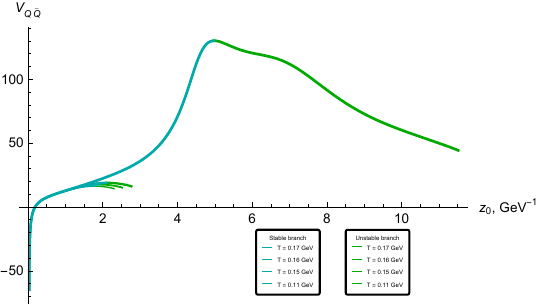}
    \subcaption{D}
\end{subfigure}

\caption{Quark-antiquark potential as a function of depth of string in the bulk for the minimal dilaton potential at temperatures $T = 0.17 \text{ GeV}, \, 0.16\text{ GeV}, \, 0.15 \text{ GeV}, \, 0.11 \text{ GeV}$ for fixed $\mu = 0 \text{ GeV (A)}, \, \mu = 0.048 \text{ GeV (B)}, \,\mu = 0.15 \text{ GeV (C)}, \,\mu = 0.3 \text{ GeV (D)}$.}
 \label{Fig:QQVzfixmu}
\end{figure}

\begin{figure}[h!]
  \centering
 $\mu=0$ GeV\hspace{150pt} $\mu=0.048$ GeV\\
 \begin{subfigure}{0.45 \textwidth}
     \includegraphics[scale=0.9]{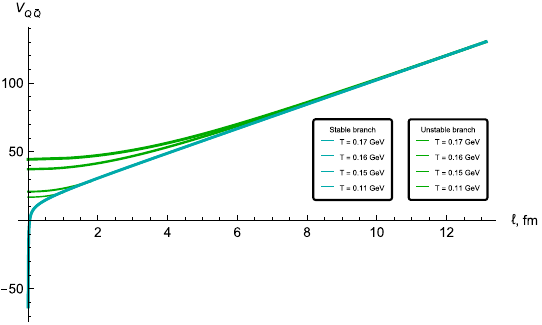}
     \subcaption{A}
 \end{subfigure}
\hfill
\begin{subfigure}{0.45 \textwidth}
    \includegraphics[scale=0.9]{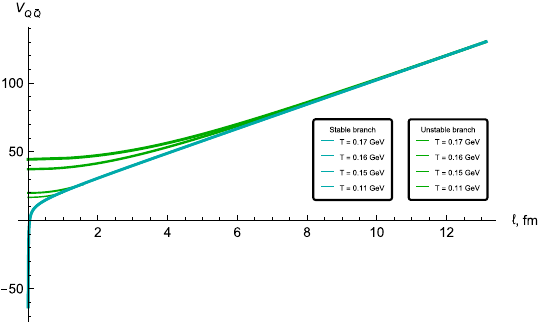}
    \subcaption{B}
\end{subfigure}\\\,\\
 $\mu=0.15$ GeV\hspace{150pt} $\mu=0.3$ GeV\\
 \begin{subfigure}{0.45 \textwidth}
     \includegraphics[scale=0.9]{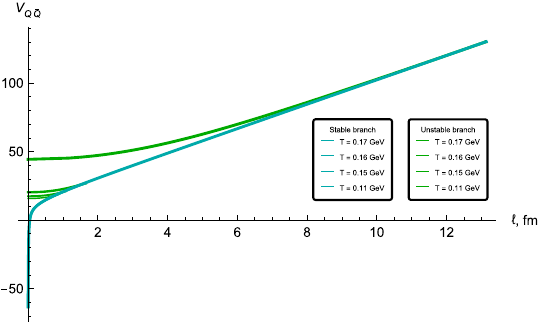}
     \subcaption{C}
 \end{subfigure}
\hfill
\begin{subfigure}{0.45 \textwidth}
    \includegraphics[scale=0.9]{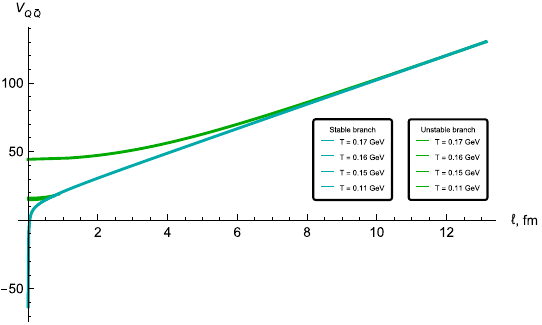}
    \subcaption{D}
\end{subfigure}

\caption{Quark-antiquark potential for the minimal dilaton potential at temperatures $T = 0.17 \text{ GeV}, \, 0.16 \text{ GeV}, \, 0.15 \text{ GeV}, \, 0.11$ GeV for fixed $\mu = 0 \text{ GeV (A)}, \, \mu = 0.048 \text{ GeV (B)}, \,\mu = 0.15 \text{ GeV (C)}, \,\mu = 0.3 \text{ GeV (D)}$.}
 \label{Fig:QQfixmu}
\end{figure}

The dependencies of the distance between quark and antiquark $\ell$ and the quark-antiquark potential $V_{Q\bar{Q}}$ over the string bulk depth $z_0$ for the direct method with the minimal dilaton potential are depicted in Fig. \ref{Fig:QQLzfixmu} and \ref{Fig:QQVzfixmu}. The corresponding dependence of quark-antiquark potential $V_{Q\bar{Q}}$ over the distance between quark and antiquark $\ell$ in different phases is depicted in Fig. \ref{Fig:QQfixmu}.  The $\ell(z_0)$ and $V_{Q\bar{Q}}(z_0)$
dependencies have a maximum point before the horizon and decline after reaching
it, while the potential $V_{Q\bar{Q}}(\ell)$ is a double-valued function. We label the branch with
the lesser potential $V_{Q\bar{Q}}$ at fixed $\ell$ as the ”stable branch”, while the other is labeled
as the ”unstable branch”. The stable branch of the potential exhibits a linear growth at
large $\ell$, but it bounces back via the unstable branch after reaching a maximum
point. Different values of chemical potential $\mu$ or temperature $T$ correspond to different maximum points and lower $\ell$ and $V_{Q\bar{Q}}$
values overall. We note that two branches were observed for all values of temperature and chemical potential.

\clearpage

\subsection{The resulting phase diagram} \label{ss:phasedg}

\begin{figure}[h!]
 \centering
 \begin{subfigure}{0.8 \textwidth}

   \includegraphics[scale=1.3]{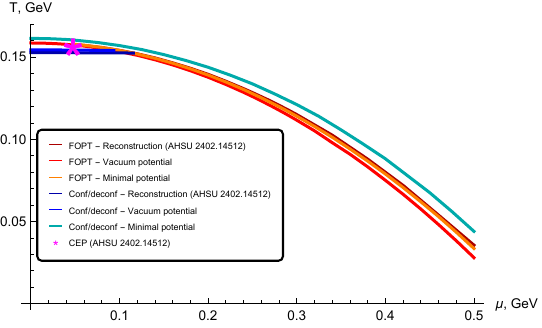} 
    \subcaption{A}
\end{subfigure}

\begin{subfigure}{0.8 \textwidth}

   \includegraphics[scale=1.3]{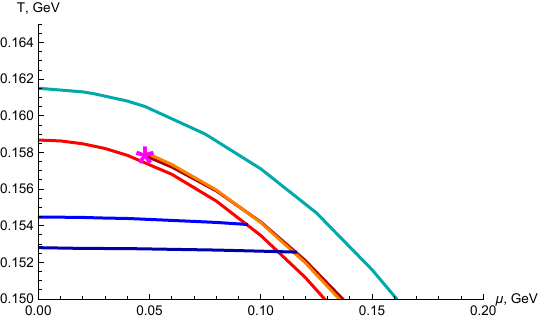} 
    \subcaption{B}
\end{subfigure}

\caption{A) HQCD phase diagram  obtained by different methods. Phase transition lines are labeled as "Phase transition type--Holographic method/dilaton potential". The FOPT and confinement/deconfinement phase transition lines, obtained using the reconstruction method, are dark red and dark blue, respectively. The FOPT and confinement/deconfinement phase transition lines, obtained using the vacuum dilaton potential, are red and blue, respectively. The FOPT and confinement/deconfinement phase transition lines, obtained using the minimal dilaton potential, are orange and cyan, respectively. Here AHSU 2402.14512 means \cite{Arefeva:2024vom}. B) is the zoom of A near the CEP.}
 \label{Fig:PD}
\end{figure}

\qquad The resulting phase diagram for the theories with potentials $\cV_1(\varphi)$, $\cV_2(\varphi)$ and the original reconstruction theory is depicted in Fig. \ref{Fig:PD} on larger and smaller scales, with $\mu = 0.5$ GeV being the largest considered chemical potential. 

The FOPT lines obtained with different methods are remarkably similar for $\mu > 0.048$ GeV beyond the CEP position. 

The confinement/deconfinement crossover lines are shown only up to their intersections with the corresponding FOPT lines. For the minimal-potential model $\cV_2(\phi)$, the cyan curve is the crossover line defined by the
finite-distance criterion of Section \ref{ss:min}. The crossover lines obtained in the reconstruction method and in the direct method with the vacuum potential are close to one another. Within the adopted finite-distance criterion, the region between the FOPT line and the cyan crossover line in the minimal-potential model may be interpreted as a quarkyonic phase.
\clearpage

\section{Conclusion and discussion} \label{s:con}

\qquad In this work, we explored two distinct methods for bottom-up holographic models, namely, the potential reconstruction method and the direct method. The potential reconstruction method starts with the known form of some of the solutions describing the black-hole background  (namely, the warp-factor $A(z)$ and the coupling function $\ff_0(z)$) and proceeds to evaluate step by step all other parts of the solution and "reconstruct" the dilaton potential. The direct method starts from the known form of the coupling function $\ff_0(\varphi)$ and the dilaton potential $\cV(\varphi)$ and proceeds to solve the EOM of the theory.

Our attempt to use the direct method to reproduce the results from the light quarks model \cite{Li:2017tdz, Arefeva:2020byn} yielded some unexpected findings. As it turned out, the coefficients of the dilaton potential and the resulting properties, such as the phase diagram of the theory, rely greatly on the fitting algorithm and the specific point used to obtain the $\cV(\varphi)$ dependence. We have described two ways to obtain the potentials for the direct method and compared the results. Each of the theories successfully reproduces one of the phase transitions but fails to reproduce the second one presented in the light quarks model. For the model with the minimal dilaton potential, we introduced a
phenomenological finite-distance criterion for the crossover based on the Cornell-like behavior of the quark--antiquark potential up to $\ell=1.5\,\mathrm{fm}$. This criterion yields a crossover line that, together with the FOPT line, delineates an intermediate region which may be interpreted as a quarkyonic phase.

Several directions for further research naturally follow from this work.  First, the behavior of other thermodynamical quantities, the running coupling constant, the jet quenching coefficient and the drag force deserves further investigation, especially across the phase transition line. Second, it
would be interesting to explore other potential reconstruction models and try to reproduce their results in the direct method to investigate more realistic models and to understand whether the discrepancy between the phase transitions is a feature of a given model or of the method itself; this warrants further exploration. Third, both the direct and reconstruction methods may be tested on models with known analytical solutions, like \cite{Arefeva:2018jyu}.

\section*{Acknowledgements} 
This work was performed within the framework of the state assignment at Steklov Mathematical Institute of Russian Academy of Sciences.

\appendix
\newpage

\section{EOM for the used holographic coordinates} \label{App:EOM}

\qquad Equations of motion (EOM) \eqref{EOMggen}-\eqref{EOMAgen} can be obtained both for the conformal coordinate ansatz \eqref{metric_z}
\bea
\label{phi2prime}
\varphi''+\left(\frac{g'}{g}+3\mathcal{A}'-\frac{3}{z}\right)\varphi'+\left(\frac{z^2e^{-2\mathcal{A}}\left(A'_t\right)^2\ff_{0;\varphi}}{2g}-\frac{e^{2\mathcal{A}}V_{\varphi}}{z^2g}\right)=0,\\\label{At2prime}
A''_t+\left(\frac{\ff_{0;\varphi}\varphi'}{\ff_{0}}+\mathcal{A}'-\frac{1}{z}\right)A'_t=0,\\
\label{phiprime}
\mathcal{A}''-\mathcal{A}'^2+\frac{2}{z}\mathcal{A}'+\frac{\varphi'^2}{6}=0,\\\label{g2prime}
g''+\left(3\mathcal{A}'-\frac{3}{z}\right)g'-e^{-2\mathcal{A}}z^2\ff_0 A_t'^2=0,\\\label{A2primes}
\mathcal{A}''+3\mathcal{A}'^2+\left(\frac{3g'}{2g}-\frac{6}{z}\right)\mathcal{A}'-\frac{1}{z}\left(\frac{3g'}{2g}-\frac{4}{z}\right)+\frac{g''}{6g}+\frac{e^{2\mathcal{A}}\cV}{3z^2g}=0,
\eea
and for the domain wall coordinate \eqref{mdw}
\bea \label{phi2}
    \varphi'' + \varphi' \left(4 A' + \cfrac{g'}{g} \right)  + \cfrac{e^{-2 A} A_t'^2 \ff_{0;\varphi} - 2 \cV_{\varphi}}{2 g} = 0,\\
    \label{At2} A_t'' + 2 A' A_t' + \cfrac{A_t' \ff_{0;\varphi} \varphi'}{\ff_0} = 0,\\
    \label{g2} g'' + 4 A' g' - e^{-2 A} \ff_0 A_t'^2 = 0,\\
    \label{A2} 6A'' + \varphi'^2 = 0,\\
    \label{constraint} 2 \cV + e^{-2 A } \ff_0 A_t'^2 + 6 A' (4 g A' + g') -g \varphi'^2 = 0.
\eea 
In both cases, the last equation is a constraint; notation $'$ means derivative over the used holographic coordinate, while $_{\varphi}$ means derivative over the dilaton field.

\newpage

\section{Details of the shooting calculation} \label{App:shoot}

\subsection{The initial value problem} \label{App:IVP}

\qquad To start the initial value problem (IVP), we need the values of the functions at the horizon and the horizon position. The total amount of unknowns and constraints at the horizon are listedd in Table \ref{t:shoot}.
\begin{table}[h!] 
\begin{center}
    \begin{tabular}{|c|c|c|c|c|c|c|c|}
    \hline
      \multicolumn{3}{|c|}{Total unknowns} & \multicolumn{4}{c|}{Total constraints} &
      \makecell{Total}\\
     \hline
     \makecell{Function \\ values} &  \makecell{Horizon \\ position}&  \makecell{Unknown \\ total}& \makecell{Horizon \\ conditions}& \makecell{Regularity \\ of equations}& \makecell{ Symmetries\\ }& 
     \makecell{ Constraints\\  total}&
     \makecell{DOF}\\
     \hline 
      8 & 1 & 9 & 2 &2 & 3 &7&2\\
      \hline 
\end{tabular}
\end{center}
\caption{Total unknowns and equations for the shooting method.}
\label{t:shoot}
\end{table}

It can be shown that the EOM \eqref{phi2}-\eqref{constraint} possess symmetries which also keep the quantities $A_\mu d x^{\mu}, \ ds^2$ constant:
\bea \label{trans}
 r \rightarrow r + a, \\
\label{times} r \rightarrow b r, \qquad g \rightarrow b^2 g, \qquad A_t \rightarrow b A_t, \qquad t \rightarrow b^{-1} t,\\
\label{spaces} A \rightarrow A + \ln(c), \qquad A_t \rightarrow c A_t, \qquad t \rightarrow c^{-1} t, \qquad \vec{x} \rightarrow c^{-1} \vec{x}.
\eea

Regularity of equation \eqref{phi2} at the horizon leads to an additional constraint:
\bea \label{c1}
    \left(\varphi' g' - \cV_{\varphi} + \cfrac{1}{2}e^{-2 A}A_t'^2 \ff_{0;\varphi}\right)|_{r = r_h}=0. 
\eea

The constraint equation \eqref{constraint} also leads to a constraint on field values on the horizon
\bea \label{c2}
    \left(2 \cV + e^{-2 A}\ff_0 A_t'^2 + 6 A' g'\right)|_{r=r_h} = 0.
\eea

Keeping in mind the conditions \eqref{horcon1}-\eqref{horcon2}, the symmetries \eqref{trans}-\eqref{spaces} and the constraints \eqref{c1}-\eqref{c2}, we arrive at a situation described in Table \ref{t:shoot}, with two remaining degrees of freedom. It is convenient to choose them as the dilaton value at the horizon $\varphi_h$ and the derivative of the vector field $a_1$ at the horizon.

We use the symmetries \eqref{trans}-\eqref{spaces} to rescale the coordinate and fields as follows:
\bea
    r_h = 0, \label{rscale1}\  A(r_h) = 0,\label{rscale2}\  g'(r_h) = 1.\label{rscale}
\eea

To deal with the $\frac{1}{g}$ terms in the equations, we start not at $r = r_h = 0$ but rather at a small offset from the horizon $r = \epsilon$. To recalculate all function values at the horizon, we expand all functions in a power series
\bea
    f(r) = \sum_{n=0}^{\infty} r^n f_n,
\eea
truncate the series at some order $N$, and calculate all coefficients from the equations \eqref{phi2}-\eqref{A2} (a more detailed derivation can be found in Appendix \ref{App:B}).

The IVP solution proceeds up to the point close to the boundary, where the functions reach their asymptotic behavior. It is mostly controlled by the UV region (where $\varphi \to 0$) asymptotics of the potential, for which we assume the form
\bea \label{UV-V}
    \cV(\varphi) = -12 + \cfrac{1}{2}m^2 \varphi^2 + \cO \left(\varphi^3\right). 
\eea

Here the dilaton mass $m^2$ is connected to the scaling dimension of the dilaton as 
\bea
    m^2 = \Delta(\Delta-4), \ \Delta = \Delta_{\pm}(\text{2 values)},\\
    \Delta_{\pm} = 2 \pm \sqrt{4 + m^2},
 \eea
 where we label $\upnu = \Delta_-$ and $\Delta = \Delta_+$. We consider masses $-4 < m^2 <0$ above the Breitenlohner--Freedman bound \cite{Breitenlohner:1982jf}, corresponding to $0 < \upnu < 2 < \Delta < 4$.

It can be shown that solutions corresponding to the asymptotically AdS space at $r \to \infty$ with boundary conditions $\varphi(\infty) = 0, \ A_t(\infty) = \mu, \ g(\infty) = 1$ have leading-order asymptotics (see App. \ref{App:C} for more details):
\bea
    \label{as_true_phi1} \varphi(r) = \varphi_A e^{ - \upnu r}  + \cO\left(e^{-2\upnu r}\right)\text{ if }\upnu<\cfrac{4}{3}, \\
    \label{as_true_phi2} \varphi(r) = \varphi_A e^{ - \upnu r}+   \cO\left(e^{-\Delta  r}\right)\text{ if }\upnu\ge\cfrac{4}{3}, \\
\label{as_true_At} A_t(r) = \mu +  \cO\left(e^{-2 r}\right),\\
    \label{as_true_g} g(r) = 1 + \cO(e^{-4r}),\\
    \label{as_true_A} A(r) = r+ \cO\left(e^{-2\upnu r}\right).
\eea
We will consider the case $\upnu < \frac{4}{3}$.

An important parameter is the characteristic energy scale$(_{es})$ of the theory, defined using the asymptotics \eqref{as_true_phi1}-\eqref{as_true_A}:
\bea
    \Lambda_{es} = \varphi_A^{\frac{1}{\upnu}}.
\eea

Since $A''<0$ from \eqref{A2}, while $A(r)$ in \eqref{as_true_A} grows asymptotically at $r\to \infty$, we introduce a constraint $A'>0$. Using formulas from Appendix \ref{App:B}, we can rewrite it as a constraint on values of $a_1$:
\bea \label{lim1}
 a_1 \le a_{c1} = \sqrt{\cfrac{-2 \cV(\varphi_h)}{\ff_0(\varphi_h)}}.
\eea

If we consider solutions with a monotonic dilaton field, we get another constraint from App. \ref{App:B}:
\bea \label{lim2}
    a_1 \le a_{c2} = \sqrt{\cfrac{2 \cV'(\varphi_h)}{\ff_0'(\varphi_h)}}.
\eea

We constraine $a_1$ by the minimum of these values \eqref{lim1}-\eqref{lim2}:
\bea
    a_1 \le \mathrm{Min}(a_{c1},a_{c2}).
\eea

Asymptotics \eqref{as_true_phi1}-\eqref{as_true_A} transform under the symmetries \eqref{trans}-\eqref{spaces} into the following:
\bea
    \label{as_phi} \varphi(r) = \tilde{\varphi}_A e^{- \upnu \alpha(r)}+\cO \left(e^{-\Delta \alpha(r)}\right),\\
    \label{as_At} A_t(r) = \tilde{A}_{t0} +\cO \left( e^{-2 \alpha(r)}\right),\\
    \label{as_g} g(r) = \tilde{g}_0 +\cO \left( e^{-4 \alpha(r)}\right),\\
    \label{as_A} A(r) = \alpha(r) + \cO \left( e^{- 2\upnu \alpha(r)}\right),
\eea
where we introduce the quantity representing the non-zero asymptotic warp factor part
\bea \label{alpha}
    \alpha(r) = \tilde{A}_{-1}r + \tilde{A}_0.
\eea

The coefficients $\tilde{A}_0, \ \tilde{A}_{-1}, \ \tilde{A}_{t0}, \ \tilde{g}_0$ can be extracted with great precision from the numerical asymptotics. Extracting $\tilde{\varphi}_A$ is associated with greater numerical noise, so in order to do that, following \cite{Critelli:2017oub},  we solve the EOM numerically until reaching two small values of the dilaton $\varphi_1 > \varphi_2$ and find corresponding $r_1 < r_2$ such that:
\bea
    \varphi(r_1) = \varphi_1, \\
    \varphi(r_2) = \varphi_2.
\eea
We extract $\tilde{A}_0, \ \tilde{A}_{-1}, \ \tilde{A}_{t0}$ and $ \tilde{g}_0$ from the field values at $r = r_1$, and approximate the dilaton as $\varphi(r) = \tilde{\varphi}_A e^{-\upnu\alpha(r)}$ to extract $\tilde{\varphi}_A$. The improved version of Fig. \ref{Fig:r-axis0} thus takes the form:
\begin{figure}[h!]
  \centering
  \includegraphics[scale=0.7]{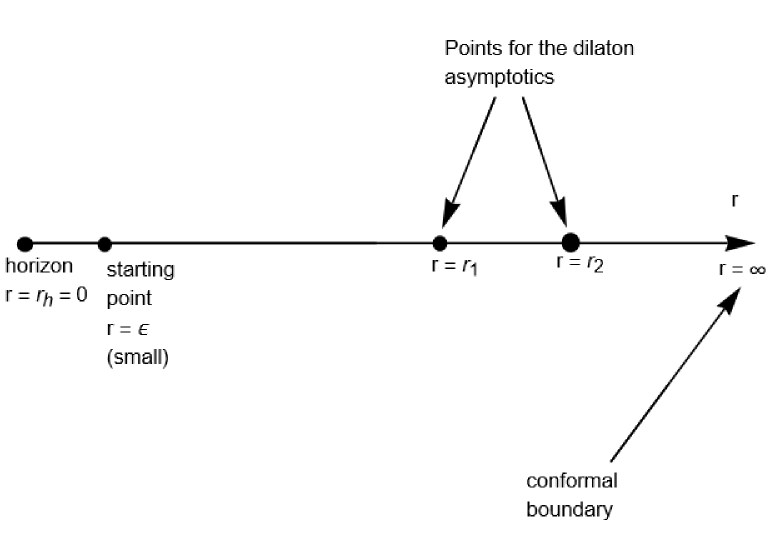} 
\caption{Coordinate axis showing points that are important for calculations.}
 \label{Fig:r-axis}
\end{figure}

Afterwards, we use symmetries \eqref{trans}-\eqref{spaces} to rescale the fields and transform asymptotics \eqref{as_phi}-\eqref{as_A} back to \eqref{as_true_phi1}-\eqref{as_true_A}. Transformations for the fields and coordinates, with subscript $_{true}$ indicating fields with correct asymptotics while its abscence indicates "numerical" fields from calculations, are as follows:
\bea
 \label{true_coords} t_{true} = t \cfrac{\sqrt{\tilde{g}_0} }{\tilde{\Lambda}},\qquad
 \vec{x}_{true} = \cfrac{\vec{x}}{\tilde{\Lambda}},\qquad
r_{true} = \tilde{A}_{-1} r + \tilde{A}_0 + \ln{\tilde{\Lambda}},\\
\label{true_fields}  \varphi_{true}(r_{true}) = \varphi(r),\qquad A_{t, true}(r_{true}) = \tilde{\Lambda} \cfrac{A_t(r)}{\sqrt{\tilde{g}_0}},\\
\label{true_metric}\qquad A_{true}(r_{true}) = A(r) + \ln(\tilde{\Lambda}), \qquad g_{true}(r_{true}) = \cfrac{g(r)}{\tilde{g}_0},
\eea
where we also introduce a numerical energy scale $\tilde{\Lambda} = \frac{\Lambda_{es}}{\tilde{\varphi}_A^{\frac{1}{\upnu}}}$.

Thermodynamic quantities can be calculated according to \eqref{chempot}-\eqref{entropy}. The formulas use the coefficients from \eqref{as_phi}-\eqref{as_A} and take the form:
\bea
    \mu = \cfrac{\tilde{A}_{t,0}\tilde{\Lambda}}{\sqrt{\tilde{g}_0}},\
    T = \cfrac{\tilde{\Lambda}}{4 \pi \sqrt{\tilde{g}_0}},\
    S = \cfrac{\tilde{\Lambda}^3}{4 G_5}.
\eea

\subsection{Series expansion of functions near the horizon} \label{App:B}

\qquad We expand all functions $\{ \varphi(r), \ A_t(r), \ g(r), \ A(r)\}$ near the horizon $r = r_h = 0$ as
\bea 
    f(r) = \sum_{n=0}^{\infty} f_n r^n.
\eea

Using the conditions \eqref{horcon1}-\eqref{horcon2} and the rescaling results \eqref{rscale}, we can write the expansion for each function:
\bea
    \varphi(r) = \varphi_h + \sum_{n=1}^{\infty} \varphi_n r^n, \label{expf}\\
    A_t(r) = a_1 r + \sum_{n=2}^{\infty} A_{t,n} r^n,\label{expt}\\
    g(r) = r + \sum_{n=2}^{\infty}g_n r^n,\label{expg}\\
    A(r) = \sum_{n=1}^{\infty}A_n r^n. \label{expA}
\eea

We substitute the series \eqref{expf}-\eqref{expA} into the equations \eqref{phi2}-\eqref{constraint} and require the EOM to be satisfied at all orders in $r$. This allows us to determine all the coefficients up to any given order and to reproduce the results from \eqref{c1}-\eqref{c2}. As an example, let us list here results for all coefficients up to $N=2$ (here $f_h$ stands for the value $f(r_h)$):
\bea
     A_1 &=& \frac{-1}{6}(2 \cV_h + \ff_{0,h}a_1^2), \quad
    \varphi_1 = \cfrac{-a_1^2 \ff_{0,h}' +2 \cV'_h}{2},\nonumber\\
    g_2 &=& -4 A_1 +a_1^2 \ff_{0,h},\quad
    A_{t,2} = -\left( 2 A_1 a_1  + \cfrac{a_1 \varphi_1 \ff_{0,h}'}{\ff_{0,h}}\right), \quad
    A_2 = -\cfrac{\varphi_1^2}{6},\nonumber\\
    \varphi_2 &=& 2 A_1 \varphi_1 + \cfrac{g_2 \varphi_1}{4} - \cfrac{A_1 a_1^2 \ff_{0,h}'}{2} + \cfrac{a_1 A_{t,2}\ff_{0,h}'}{2} - \cfrac{a_1^2 g_2 \ff_{0,h}'}{8} + \cfrac{g_2 \cV_h'}{4}+\cfrac{a_1^2 \varphi_1 \ff_{0,h}''}{4} - \cfrac{\varphi_1 \cV_h''}{2}.\nonumber
\eea

\subsection{Derivation of asymptotics near the conformal boundary} \label{App:C}

\qquad Let us obtain the leading-orders asymptotic solutions for the EOM system \eqref{phi2}-\eqref{A2} in the UV region  $r \to \infty$. We write asymptotic solutions for functions (labeled $f(r)$) in the form
\bea
    f(r) = f_{as(1)}(r) +  f_{as(2)}(r) + \dots \text{ for } r \to \infty, 
\eea
where $f_{as(i)}(r)\lll f_{as(j)}(r)$ for $i<j$ as $r \to \infty$.

1) We consider the space near the boundary to be close to AdS. That means that the asymptotic form of the warp factor is
\bea \label{leadA}
    A_{as(1)}(r) =r,\text{ for }r \to \infty.
\eea

\begin{table}[h!]
\begin{center}
	\begin{tabular}{|c|c|c|}
		\hline
		& $0 < \upnu < 2$  \\
		\hline
		$ A_{as(1)}(r)$ & $r$ \\
		\hline
	\end{tabular}
\end{center}
\caption{Leading-order asymptotics for the warp factor.}
\label{table_A_1}
\end{table}

2) From \eqref{A2}, that is, $6A'' + \varphi'^2 = 0$, it follows that 
\bea \label{leadphi0}
   \varphi_{as(1)}'(r) \to 0\text{ for } r \to \infty.
\eea

3) From \eqref{At2}, i.e., $A_t'' + 2 A' A_t' + \cfrac{A_t' \ff_{0;\varphi} \varphi'}{\ff_0} = 0$, and using \eqref{leadA} and \eqref{leadphi0}, and assuming that $\cfrac{f_{0;\varphi}}{f_0}|_{\varphi \to 0} \ne \infty$, we get a simplified equation for the asymptotics of $A_t$:
\bea \label{leadAteq}
    A_{t,as(1)}'' + 2 A_{t,as(1)}' = 0.
\eea

This equation has a solution
\bea \label{leadAt}
    A_{t,as(1)}(r) = A_{t0} + A_{t2} e^{-2r}, 
\eea
where $A_{t0}, \  A_{t2}$ are some constants.

\begin{table}[h!]
\begin{center}
	\begin{tabular}{|c|c|c|}
		\hline
		& $0 < \upnu < 2$  \\
		\hline
		$  A_{t,as(1)}(r)$ & $A_{t0} + A_{t2} e^{-2r}$ \\
		\hline
	\end{tabular}
\end{center}
\caption{Leading-order asymptotics for the vector field.}
\label{table_At_1}
\end{table}

4) Asymptotics for the blackening function can be derived from \eqref{g2}, that is, $g'' + 4 A' g' - e^{-2 A} \ff_0 A_t'^2 = 0$. 

The last term in this equation
\bea
e^{-2 A} \ff_0 A_t'^2.
\eea
decreases exponentially to zero if $\ff_0(\varphi)|_{\varphi \to 0} < \infty$, so we get a simplified equation for asymptotics $g_{as}(r) $ of $g(r)$
\bea \label{leadgeq}
   g_{as(1)}'' + 4 g_{as(1)}' = 0.
\eea
Assuming that $g(r)|_{r \to \infty} = 1$, we get a solution for this equation
\bea \label{leadg}
    g_{as(1)}(r) = 1 +g_{4} e^{-4r}, 
\eea
where $g_4$ is some constant.

\begin{table}[h!]
\begin{center}
	\begin{tabular}{|c|c|c|}
		\hline
		& $0 < \upnu < 2$  \\
		\hline
		$ g_{as(1)}(r)$ & $1 + g_4 e^{ - 4 r}$ \\
		\hline
	\end{tabular}
\end{center}
\caption{Leading-order asymptotics for the blackening function.}
\label{table_g_1}
\end{table}

5) Next, we substitute into \eqref{phi2}, that is $$  \varphi'' + \varphi' \left(4 A' + \cfrac{g'}{g} \right)  + \cfrac{e^{-2 A} A_t'^2 \ff_{0;\varphi} - 2 \cV_{\varphi}}{2 g} = 0,$$ asymptotics for the potential \eqref{UV-V}, the warp-factor \eqref{leadA}, the blackening function \eqref{leadg} and the vector potential \eqref{leadAt}. Thus we obtain the equation for the dilaton asymptotics (dropping all exponentially decreasing terms):
\bea \label{leadphieq}
    \varphi_{as(1)}'' + 4\varphi_{as(1)}' - m^2 \varphi_{as(1)} = 0. 
\eea
Let us search for the solutions of this equation in the form 
\bea
    \varphi_{as(1)}(r) = K e^{-cr},
\eea
where $K$, $c$ are constants.
In that case we get an equation for $c$
\bea
    c^2 - 4 c - m^2 = 0,
\eea
for which we have solutions $c = \Delta$ and $c = \upnu$. The value of $K$ is not fixed.

The corresponding dilaton asymptotics are
\bea \label{leadphi}
    \varphi_{as(1)}(r) =  \varphi_A e^{ - \upnu r}  + \varphi_B e^{ - \Delta r},
\eea
where $\varphi_A$ and $\varphi_B$ are some constants.

\begin{table}[h!]
\begin{center}
	\begin{tabular}{|c|c|c|}
		\hline
		& $0 < \upnu < 2$  \\
		\hline
		$ \varphi_{as(1)}(r)$ & $\varphi_A e^{ - \upnu r}$ \\
		\hline
	\end{tabular}
\end{center}
\caption{Leading-order asymptotics for the dilaton field.}
\label{table_phi_1}
\end{table}

6) Combination of these asymptotics with similar analysis of the subleading terms yields the following asymptotic expansions:
\bea
    \label{A:as_true_phi1} \varphi(r) = \varphi_A e^{ - \upnu r}  + \cO\left(e^{-2\upnu r}\right)\text{ if }\upnu<\cfrac{4}{3}, \\
    \label{A:as_true_phi2} \varphi(r) = \varphi_A e^{ - \upnu r}+   \cO\left(e^{-\Delta  r}\right)\text{ if }\upnu\ge\cfrac{4}{3}, \\
\label{A:as_true_At} A_t(r) = A_{t0} + A_{t2} e^{ - 2 r}+ \cO\left(e^{-(2+\upnu) r}\right),\\
    \label{A:as_true_g} g(r) = 1 + g_4 e^{ - 4 r} + \cO\left(e^{-(4+2\upnu) r}\right) \text{ if }\upnu<1,\\
    g(r) = 1 + g_4 e^{ - 4 r} + \cO\left(e^{-6 r}\right) \text{ if } \upnu \ge 1,\\
    \label{A:as_true_A} A(r) = r+ \cO\left(e^{-2\upnu r}\right).
\eea

\newpage

\section{Dilaton potentials and coupling functions} \label{App:D}

\qquad The dilaton potentials used in the direct approach are
\bea
  \cV_1(\varphi) = -12 \cosh(0.76657 \varphi) + 2.0258 \varphi^2 +0.0893 \varphi^4 + 0.0014 \varphi^6 \nonumber\\
     + 0.0000115\varphi^8 
    + 5.79\times10^{-8}  \varphi^{10} +1.87 \times 10^{-10} \varphi^{12} \nonumber\\
+5.18 \times 10^{-13} \varphi^{14} +5.00 \times 10^{-16} \varphi^{16} +1.76 \times 10^{-18} \varphi^{18},\nonumber\\
    \cV_2(\varphi) = - 12 \cosh(0.76775 \varphi) +2.0366  \varphi^2 + 0.0715\varphi^4 + 0.0027\varphi^6 \nonumber\\
    - 6.9511 \times 10^{-6} \varphi^8 + 1.81 \times 10^{-7}\varphi^{10} 
+ 4.25 \times 10^{-10} \varphi^{12}  \nonumber\\
+1.28 \times 10^{-12} \varphi^{14} 
-4.24 \times 10^{-14} \varphi^{16} +1.63 \times 10^{-16} \varphi^{18} \nonumber.
\eea

The coupling function used in the direct approach is
\bea
 \ff_0(\varphi) = \exp(-0.068 \varphi^2 + 0.0000185 \varphi^4 - 4.5967 \times 10^{-7} \varphi^6  +1.666 \times 10^{-9} \varphi^8  \nonumber\\ - 3.841 \times 10^{-12} \varphi^{10} +5.246 \times 10^{-15} \varphi^{12} - 3.892\times 10^{-18} \varphi^{14} + 1.206 \times 10^{-21} \varphi^{16}).\nonumber
\eea


\newpage

\begin{thebibliography}{99}

\bibitem{Blankenbecler:1981jt}
R. Blankenbecler, D. J. Scalapino, and R. L. Sugar,
``Monte Carlo Calculations of Coupled Boson - Fermion Systems. 1.,''
Phys. Rev. D \textbf{24}, 2278 (1981).

\bibitem{Maldacena:1997re}
J. M. Maldacena,
``The Large $N$ limit of superconformal field theories and supergravity,''
Adv. Theor. Math. Phys. \textbf{2}, 231 (1998)
[arXiv:hep-th/9711200].

\bibitem{Witten:1998zw}
E. Witten,
``Anti-de Sitter space, thermal phase transition, and confinement in gauge theories,''
Adv. Theor. Math. Phys. \textbf{2}, 505 (1998)
[arXiv:hep-th/9803131].

\bibitem{Karch:2002sh}
A. Karch and E. Katz,
``Adding flavor to AdS / CFT,''
JHEP \textbf{06}, 043 (2002)
[arXiv:hep-th/0205236].

\bibitem{Sakai:2004cn}
T. Sakai and S. Sugimoto,
``Low energy hadron physics in holographic QCD,''
Prog. Theor. Phys. \textbf{113}, 843 (2005)
[arXiv:hep-th/0412141].

\bibitem{Erlich:2005qh}
J. Erlich, E. Katz, D. T. Son, and M. A. Stephanov,
``QCD and a holographic model of hadrons,''
Phys. Rev. Lett. \textbf{95}, 261602 (2005)
[arXiv:hep-ph/0501128].

\bibitem{Gursoy:2007cb}
U. Gursoy and E. Kiritsis,
``Exploring improved holographic theories for QCD: Part I,''
JHEP \textbf{02}, 032 (2008)
[arXiv:0707.1324].

\bibitem{Gursoy:2007er}
U. Gursoy, E. Kiritsis, and F. Nitti,
``Exploring improved holographic theories for QCD: Part II,''
JHEP \textbf{02}, 019 (2008)
[arXiv:0707.1349].

\bibitem{DeWolfe:2013cua}
O.~DeWolfe, S.~S.~Gubser, C.~Rosen and D.~Teaney,
''Heavy ions and string theory,''
Prog. Part. Nucl. Phys. \textbf{75}, 86-132 (2014)
doi:10.1016/j.ppnp.2013.11.001
[arXiv:1304.7794 [hep-th]].

\bibitem{DeWolfe:1999cp}
O.~DeWolfe, D.~Z.~Freedman, S.~S.~Gubser and A.~Karch,
``Modeling the fifth-dimension with scalars and gravity,''
Phys. Rev. D \textbf{62}, 046008 (2000)
doi:10.1103/PhysRevD.62.046008
[arXiv:hep-th/9909134 [hep-th]].

\bibitem{Arefeva:2005mka}
I.~Y.~Aref'eva, A.~S.~Koshelev and S.~Y.~Vernov,
``Crossing of the w = -1 barrier by D3-brane dark energy model,''
Phys. Rev. D \textbf{72}, 064017 (2005)
doi:10.1103/PhysRevD.72.064017
[arXiv:astro-ph/0507067 [astro-ph]].


\bibitem{He:2010ye}
S. He, M. Huang, and Q.-S. Yan,
``Logarithmic correction in the deformed $\rm AdS_5$ model to produce the heavy quark potential and QCD beta function,''
Phys. Rev. D \textbf{83}, 045034 (2011)
[arXiv:1004.1880].

\bibitem{Li:2017tdz}
M.-W. Li, Y. Yang, and P.-H. Yuan,
``Approaching Confinement Structure for Light Quarks in a Holographic Soft Wall QCD Model,''
Phys. Rev. D \textbf{96}, 066013 (2017)
[arXiv:1703.09184].

\bibitem{Arefeva:2018hyo}
I.~Aref'eva and K.~Rannu,
JHEP \textbf{05}, 206 (2018)
doi:10.1007/JHEP05(2018)206
[arXiv:1802.05652 [hep-th]].

\bibitem{Arefeva:2020byn}
I. Ya. Aref'eva, K. Rannu, and P. Slepov,
``Holographic anisotropic model for light quarks with confinement-deconfinement phase transition,''
JHEP \textbf{06}, 090 (2021)
[arXiv:2009.05562].

\bibitem{Arefeva:2022avn}
I. Ya. Aref'eva, A. Ermakov, K. Rannu, and P. Slepov,
``Holographic model for light quarks in anisotropic hot dense QGP with external magnetic field,''
Eur. Phys. J. C \textbf{83}, 79 (2023)
[arXiv:2203.12539].

\bibitem{Arefeva:2022bhx}
I. Ya. Aref'eva, K. A. Rannu, and P. S. Slepov,
``Anisotropic solution of the holographic model of light quarks with an external magnetic field,''
Theor. Math. Phys. \textbf{210}, 363 (2022).

\bibitem{Chen:2024mmd}
X. Chen and M. Huang,
``Flavor dependent critical endpoint from holographic QCD through machine learning,''
JHEP \textbf{02}, 123 (2025)
[arXiv:2405.06179].

\bibitem{Arefeva:2024vom}
I. Ya. Aref'eva, A. Hajilou, P. Slepov, and M. Usova,
``Running coupling for holographic QCD with heavy and light quarks: Isotropic case,''
Phys. Rev. D \textbf{110}, 126009 (2024)
[arXiv:2402.14512].

\bibitem{Arefeva:2024xmg}
I. Ya. Aref'eva, A. Hajilou, A. Nikolaev, and P. Slepov,
``Holographic QCD running coupling for light quarks in strong magnetic field,''
Phys. Rev. D \textbf{110}, 086021 (2024)
[arXiv:2407.11924].

\bibitem{Arefeva:2024poq}
I. Ya. Aref'eva, A. Hajilou, P. Slepov, and M. Usova,
``Beta-function dependence on the running coupling in holographic QCD models,''
Teor. Mat. Fiz. \textbf{221}, 2132 (2024)
[arXiv:2407.14448].

\bibitem{Arefeva:2025xtz}
I. Ya. Aref'eva, A. Hajilou, P. Slepov, and M. Usova,
``Beta-functions and RG flows for holographic QCD with heavy and light quarks: Isotropic case,''
Phys. Rev. D \textbf{111}, 046013 (2025)
[arXiv:2503.09444].

\bibitem{Deng:2026aht}
Y. Deng, M. Huang, and L. Zhang,
``Holographic QCD equation of state constrained by lattice QCD: neural-ODE for probe-limit and a back-reaction test,''
arXiv:2602.21618 (2026).

\bibitem{DeWolfe:2010he}
O. DeWolfe, S. S. Gubser, and C. Rosen,
``A holographic critical point,''
Phys. Rev. D \textbf{83}, 086005 (2011)
[arXiv:1012.1864].

\bibitem{Knaute:2017opk}
J. Knaute, R. Yaresko, and B. K\"ampfer,
``Holographic QCD phase diagram with critical point from Einstein–Maxwell-dilaton dynamics,''
Phys. Lett. B \textbf{778}, 419 (2018)
[arXiv:1702.06731].

\bibitem{Critelli:2017oub}
R. Critelli, J. Noronha, J. Noronha-Hostler, I. Portillo, C. Ratti, and R. Rougemont,
``Critical point in the phase diagram of primordial quark-gluon matter from black hole physics,''
Phys. Rev. D \textbf{96}, 096026 (2017)
[arXiv:1706.00455].

\bibitem{Cai:2022omk}
R.-G. Cai, S. He, L. Li, and Y.-X. Wang,
``Probing QCD critical point and induced gravitational wave by black hole physics,''
Phys. Rev. D \textbf{106}, L121902 (2022)
[arXiv:2201.02004].

\bibitem{Jokela:2024xgz}
N. Jokela, M. J\"arvinen, and A. Piispa,
``Refining holographic models of the quark-gluon plasma,''
Phys. Rev. D \textbf{110}, 126013 (2024)
[arXiv:2405.02394].

\bibitem{Li:2025lmp}
Z. Li and F. Wang,
``Phase transition of hot dense QCD Matter from a refined holographic EMD model,''
arXiv:2507.09113 (2025).

\bibitem{Shen:2025yrn}
J.-Y. Shen, X.-Y. Liu, J.-R. Wu, Y.-L. Wu, and Z. Fang,
``Phase structure of 2+1-flavor QCD from an Einstein-dilaton-flavor holographic model,''
JHEP \textbf{06}, 030 (2026)
[arXiv:2511.19127].

\bibitem{Shen:2025zkj}
J.-Y. Shen, X.-Y. Liu, J.-R. Wu, Y.-L. Wu, and Z. Fang,
``Toward a holographic realization of the 2+1-flavor QCD phase structure,''
Phys. Rev. D \textbf{112}, L111504 (2025)
[arXiv:2511.11273].

\bibitem{Chamblin:1999tk}
A. Chamblin, R. Emparan, C. V. Johnson, and R. C. Myers,
``Charged AdS black holes and catastrophic holography,''
Phys. Rev. D \textbf{60}, 064018 (1999)
[arXiv:hep-th/9902170].

\bibitem{Hawking:1974rv}
S. W. Hawking,
``Black hole explosions,''
Nature \textbf{248}, 30 (1974).

\bibitem{Bekenstein:1973ur}
J. D. Bekenstein,
``Black holes and entropy,''
Phys. Rev. D \textbf{7}, 2333 (1973).

\bibitem{Hawking:1975vcx}
S. W. Hawking,
``Particle Creation by Black Holes,''
Commun. Math. Phys. \textbf{43}, 199 (1975)
[Erratum: \textbf{46}, 206 (1976)].

\bibitem{Wilson:1974sk}
K. G. Wilson,
``Confinement of Quarks,''
Phys. Rev. D \textbf{10}, 2445 (1974).

\bibitem{Maldacena:1998im}
J. M. Maldacena,
``Wilson loops in large N field theories,''
Phys. Rev. Lett. \textbf{80}, 4859 (1998)
[arXiv:hep-th/9803002].

\bibitem{Rey:1998ik}
S.-J. Rey and J.-T. Yee,
``Macroscopic strings as heavy quarks in large N gauge theory and anti-de Sitter supergravity,''
Eur. Phys. J. C \textbf{22}, 379 (2001)
[arXiv:hep-th/9803001].

\bibitem{Andreev:2006ct}
O. Andreev and V. I. Zakharov,
``Heavy-quark potentials and AdS/QCD,''
Phys. Rev. D \textbf{74}, 025023 (2006)
[arXiv:hep-ph/0604204].

\bibitem{PDBook}
W.-M. Yao \textit{et al.} (Particle Data Group),
``Review of Particle Physics,''
J. Phys. G \textbf{33}, 1 (2006).

\bibitem{Bazavov:2017dus}
A. Bazavov \textit{et al.},
``The QCD Equation of State to $\mathcal{O}(\mu_B^6)$ from Lattice QCD,''
Phys. Rev. D \textbf{95}, 054504 (2017)
[arXiv:1701.04325].


\bibitem{Breitenlohner:1982jf}
P. Breitenlohner and D. Z. Freedman,
``Stability in Gauged Extended Supergravity,''
Annals Phys. \textbf{144}, 249 (1982).

\bibitem{Arefeva:2018jyu}
I.~Y.~Aref'eva, A.~A.~Golubtsova and G.~Policastro,
``Exact holographic RG flows and the A$_{1}$ {\texttimes} A$_{1}$ Toda chain,''
JHEP \textbf{05}, 117 (2019)
doi:10.1007/JHEP05(2019)117
[arXiv:1803.06764 [hep-th]].

\end{thebibliography}
\end{document}